\documentclass[11pt,a4paper]{article} 
\pdfoutput=1 
\usepackage{jcappub} 
\usepackage{graphicx}
\usepackage{dcolumn}
\usepackage{bm}
\usepackage[english]{babel}
\usepackage[latin1]{inputenc} 
\usepackage{graphicx}
\usepackage{epstopdf}
\usepackage{amsfonts}
\usepackage{color}
\usepackage{comment}
\usepackage{epsfig}
\usepackage{mathrsfs}
\usepackage{amsmath,amssymb}
\usepackage{subfigure}

\usepackage{framed}
\usepackage{lipsum}
\usepackage[svgnames]{xcolor}
\definecolor{shadecolor}{named}{LightGrey}

\newcommand{\erf}{\text{erf}}
\newcommand{\lto}[1]{\longrightarrow#1}

\renewcommand{\(}{\left(}
\renewcommand{\)}{\right)}
\renewcommand{\[}{\left[}
\renewcommand{\]}{\right]}

\newcommand{\Xmax}{X_{\text{max}}}

\def\g\,cm{\ifmmode {\mathrm{g/cm}^2}\else
                   {g/cm$^2$}\fi\xspace}%
\def\sigmaXmax{\ifmmode {\mathrm{RMS}(X_\mathrm{max})}\else
                   {RMS$(X_\mathrm{max})$}\fi\xspace}%
\def\meanXmax{\ifmmode {\langle X_\mathrm{max}\rangle}\else
                   {$\langle X_\mathrm{max}\rangle$}\fi\xspace}%

\def\eV{\ifmmode {\mathrm{\ e\kern -0.1em V}}\else
                   \textrm{e\kern -0.1em V}\fi\xspace}%

\begin{document}
\selectlanguage{english}

\title{Reinterpreting the development of extensive air showers initiated by nuclei and photons} 

\author[a,b,c]{Manlio De Domenico\note{Now at Departament d'Enginyeria Inform\'atica i Matem\'atiques, Universitat Rovira i Virgili, Tarragona, Spain.},}
\author[d]{Mariangela Settimo,}
\author[e]{Simone Riggi,}
\author[f]{Eric Bertin}

\affiliation[a]{Laboratorio sui Sistemi Complessi, Scuola Superiore di Catania, \\
Via Valdisavoia 9, 95123 Catania, Italy}
\affiliation[b]{Istituto Nazionale di Fisica Nucleare, Sez. di Catania, \\
Via S. Sofia 64, 95123 Catania, Italy}
\affiliation[c]{Dipartimento di Fisica e Astronomia, Universit\'a degli Studi di Catania, \\
Via S. Sofia 64, 95123 Catania, Italy}
\affiliation[d]{Laboratoire de Physique Nucl\'eaire et de Hautes Energies (LPNHE), Universit\'es Paris 6 et Paris 7, CNRS-IN2P3, Paris, France, \\
University of Siegen, Germany}
\affiliation[e]{INAF, Osservatorio Astrofisico di Catania, Italy}
\affiliation[f]{Universit\'e de Lyon, Laboratoire de Physique,
\'Ecole Normale Sup\'erieure de Lyon, CNRS,
46 all\'ee d'Italie, F-69007 Lyon, France}

\emailAdd{manlio.dedomenico@ct.infn.it}
\emailAdd{settimo@hep.physik.uni-siegen.de}
\emailAdd{simone.riggi@ct.infn.it}
\emailAdd{eric.bertin@ens-lyon.fr}

\date{\today}
\abstract{
Ultra-high energy cosmic rays (UHECRs) interacting with the atmosphere generate extensive air showers (EAS) of secondary particles. The depth corresponding to the maximum development of the shower, $\Xmax$, is a well-known observable for determining the nature of the primary cosmic ray which initiated the cascade process.
In this paper, we present an empirical model to describe the distribution of $\Xmax$ for EAS initiated by nuclei, in the energy range from $10^{17}$~eV up to $10^{21}$~eV, and by photons, in the energy range from $10^{17}$~eV up to $10^{19.6}$~eV. Our model adopts the generalized Gumbel distribution motivated by the relationship between the generalized Gumbel statistics and the distribution of the sum of non-identically distributed variables in dissipative stochastic systems. We provide an analytical expression for describing the $\Xmax$ distribution for photons and for nuclei, and for their first two statistical moments, namely $\langle \Xmax\rangle$ and $\sigma^{2}(\Xmax)$. The impact of the hadronic interaction model is investigated in detail, even in the case of the most up-to-date models accounting for LHC observations. We also briefly discuss the differences with a more classical approach and an application to the experimental data based on information theory.
}


\notoc

\maketitle

\flushbottom


\section{Introduction}\label{IntroductionSection}%

The composition of cosmic rays at ultra-high energies is a fundamental observable to test and constrain the present theories concerning the origin, nature and production site properties of a such extreme radiation.
The measurement of the composition is experimentally based on the correlation of shower observables to the mass of the primary particle. Among them, the atmospheric depth of shower maximum ($\Xmax$) and the number of muons at ground level ($N_{\mu}$) exhibit the strongest correlation. An exhaustive review on experimental techniques over a wide energy range is available in~\cite{kampert}.

The current results from the Pierre Auger Observatory~\cite{auger_mass1, auger_mass2}, HiRes~\cite{HiRes_mass}, Telescope Array (TA)~\cite{TA_mass} and Yakutsk~\cite{Yakutsk_mass} agree at the level of their quoted systematics suggesting a light composition below 10$^{18.5}$~eV. Whereas the $\Xmax$ measurements from Auger indicate a mixed composition changing from light to heavier component at energies of about 10$^{18.5}$~eV, the HiRes and TA results is compatible with a constant composition dominated by light elements.  A detailed comparison between these experiments and a discussion on the agreement between their results is given in~\cite{WGmass2012} as a result of a joined effort between the collaborations. 
All these experiments make use of the mean and the rms of the $\Xmax$ distributions as a function of energy to get indications of the nature of the primary cosmic ray. On the other hand the interpretation of $\Xmax$ in terms of nuclear mass rely on the predicted $\Xmax$ values from Monte Carlo simulations of the shower development in atmosphere (see, for instance, \cite{augerXmaxInterpretation}).  
They are strongly based on detailed models of the hadronic interactions at very high energies and therefore require to extrapolate the interaction parameters from laboratory experiments performed at lower energies. This introduces an additional source of uncertainty in the interpretation of the composition measurements. 
In addition, the analysis of composition data typically requires to employ a large sample of shower simulations with more than one hadronic model on the market, different primary nuclei and a wide range of primary energies and angles. 

This context motivates the development of an analytical model to describe the distribution of the depth of the shower maximum parameter not only to speed-up the composition analysis based on $\Xmax$ (elongation rate, RMS($\Xmax$), $\Xmax$ distributions fitting) but also to overcome the limitations related to the low event statistics.

In this paper we introduce a method to describe the distribution of $\Xmax$ by applying statistical methods typically developed for special dissipative stochastic systems, i.e., dynamical systems of many particles governed by stochastic rules, and subjected to energy injection and dissipation, keeping the whole process far from thermodynamic equilibrium.

In the literature \cite{desouza}, the fluctuations observed in the position of the $\Xmax$, for a given primary particle and energy, can be related to the stochastic fluctuations in (i) the position of the first interaction point $X_{\rm{first}}$ in the top layers of the atmosphere and (ii) the secondary interactions occurring along the shower development.

Thus, by assuming that such interactions are not correlated to each other, the $\Xmax$ distribution can be described by the convolution of these two terms: 
\begin{align}
p_{\rm{max}}(X_{\rm{max}}) & =  \int_{0}^{\infty}{p_{1}\left(X_{\rm{max}}-\Delta \right)\,p_{\Delta}(\Delta)d\Delta} = \int_{0}^{X_{\rm{max}}}\frac{1}{\lambda_{1}}e^{-\frac{X_{\rm{max}}-\Delta}{\lambda_{1}}}\,p_{\Delta}(\Delta)d\Delta.
\label{eq:convolution}
\end{align}%
where $X_{\rm{first}}$ follows an exponential distribution $p_{1}$ with interaction length $\lambda$, $\Delta~=~\Xmax~-~X_{\rm{first}}$ refers to the shower development and $p_{\Delta}$ is its probability distribution. Since the shower development involves a large number of particles, the fluctuations in the secondary sub-cascades are assumed to be Gaussian-distributed and the $\Xmax$ distribution is parameterized with the numerical function: 
\begin{align}
\label{def-xmax-convol}
p(\Xmax)=\frac{1}{2\lambda_{1}}\exp{\( -\frac{\Xmax-\mu}{\lambda_{1}}+\frac{\sigma^{2}}{2\lambda_{1}^{2}} \)}
\[1+\erf\( \frac{\Xmax-\mu}{\sigma\sqrt{2}}-\frac{\sigma}{\lambda_{1}\sqrt{2}} \)\], 
\end{align}
where $\mu$ and $\sigma$ are the parameters of the Gaussian function, and $\erf$ is the standard error function. However, such an assumption is a rather delicate issue, because the fluctuations involve many physical processes as, for instance, the multiplicity of produced secondaries and the energy loss of the leading particle.

In this paper, we adopt an alternative approach to describe the fluctuations observed in the position of the $\Xmax$ by relating the shower development to the dynamics of special dissipative stochastic systems. In this context, it has been recently shown \cite{bertin2005global,bertin2006exactly} that the sum of correlated and/or non-identically distributed random variables in such systems follows a generalized Gumbel distribution, at least in the framework of specific models. In particular we find that this distribution describes also the distribution of $\Xmax$, thus providing the basis for a new statistical description of the shower development. Intriguingly, this result applies for any primary nucleus and photons, unrevealing the universality of the underlying process. 

In this paper we use Monte Carlo simulations of several primary particles, generated in the energy range between 10$^{17}$ and 10$^{21}$~eV and assuming different hadronic interaction models, to extract the expected $\Xmax$ distribution. Hence we fit the generalized Gumbel to the simulated $\Xmax$ and we derive the dependence of its parameters against physical observables, namely the primary energy and the primary mass. 

The paper is organized as follows. The description of our model is given in Sec.~\ref{ModelSection}. In Sec.\,\ref{ParametrizationSection} we introduce the parameterization of the generalized Gumbel function. More specifically, we provide a practical method to obtain a likely value of $\Xmax$ for all nuclei and photons, and for three different hadronic models widely adopted for such studies, i.e., \textsc{QGSjet}~II-03~\cite{qgsjet2}, \textsc{Sibyll}~2.1~\cite{sibyll} and \textsc{Epos}~1.99~\cite{epos}. Moreover, we also use the latest updated versions, including the corrections observed from LHC results~\cite{lhc_EAS}, namely \textsc{QGSjet}~II-04~\cite{qgsjet2-04} and \textsc{Epos-LHC}~\cite{epos-lhc}. Finally in Sec.~\ref{DiscussionSection} we discuss the potentiality of the proposed method.


\section{A model to describe shower maxima behavior}\label{ModelSection}%

\begin{figure}[!t]
\centering%
\includegraphics[width=10cm,clip]{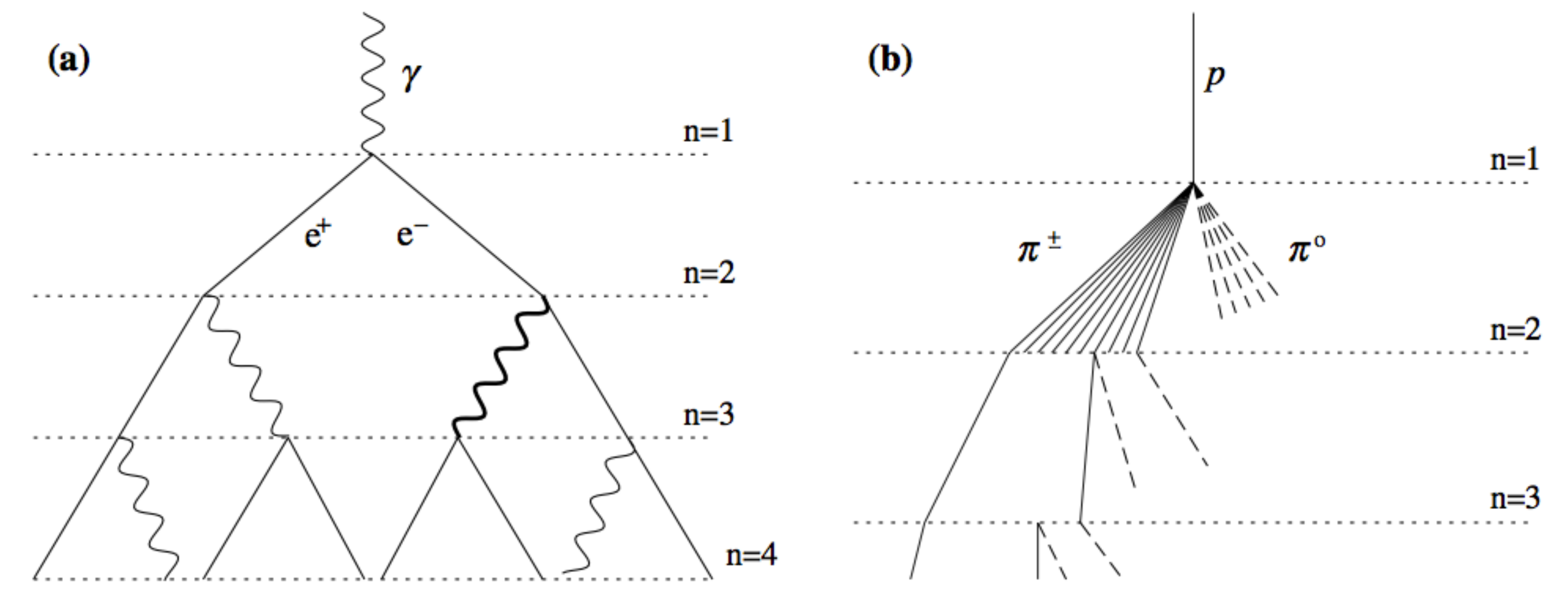}%
\caption{Schematic view of (a) electromagnetic and (b) hadronic cascades. In the latter, dashed lines represent neutral particles ($\pi^{0}$) and solid lines charged particles ($\pi^{\pm}$). Only one charged hadron interaction is shown for each generation (see Refs.\,\cite{Matthews2005,Pierog2006} for further detail). Picture from \cite{Matthews2005}.}
\label{fig-heitler}
\end{figure}

The description of the development of the EAS is not trivial because of the number of processes and correlated properties (cross sections, inelasticity, multiplicity,...) that are involved. A simplified and schematic description of the development of the particle cascade was initially proposed by Heitler~\cite{Heitler1954} for the case of the electromagnetic showers, induced by a photons or electron (positron). Unless explicitly written, hereafter we denote with electrons both electrons and its antiparticle. More recently the model has been generalized to the case of hadronic showers (i.e. induced by nuclei) in~\cite{Matthews2005,Pierog2006}. 
We refer to the original papers for a full description of the Heitler model and of its generalized version. A sketch of the two models, is illustrated in Fig.\,\ref{fig-heitler}. In the case of the electromagnetic component (Fig.\,\ref{fig-heitler}a) each particle, namely photons ($\gamma$), electrons ($e^-$) and positrons ($e^+$), traverse the atmosphere for a radiation length ($X_0$) before interacting or decaying until the energy falls below the critical energy $E_c$. Therefore, each particle produces two new particles according to the dominant processes (pair production $e^+\,e^-$ for photons and Bremsstrahlung radiation for electrons). In this model two other assumptions are done: the two produced particles equally share the parent's energy and the interaction length is the same for all particles. From this simple model, a rough estimate of the $\Xmax$ is obtained as $\Xmax~=~X_0\, \ln(E_0/E_c)$, with E$_0$ the initial energy of the particle inducing the cascade. 

In the case of the hadronic component (Fig.\,\ref{fig-heitler}b) the process involves the production of many particles of different species, the multiplicity of newly generated particles is not constant, and the interaction length plays the role of the radiation length. However, the development of the hadronic component shares with the electromagnetic case the same general structure, although different physical interactions are involved.

In this section we investigate the basic Heitler model and its generalized counterpart from the point of view of statistics.
The Heitler model, with its assumptions, is a simplification valid only for limited aims and contexts. Indeed, with no regards for the particle type, it is more realistic to assume that, at each step of the shower development, the probability of interaction (or decay) per unit length follows the exponential law $f(x;\lambda)=\lambda^{-1}\exp(-x/\lambda)$, where $\lambda$ is the mean free path. On the other hand, the number of secondary particles rapidly increases with the atmospherical depth until the critical energy is reached. The presence of nuclear particles of different species makes difficult an exact description of the statistics of their number.
However, to obtain $\Xmax$, we are interested in the atmospherical depth corresponding to the maximum number $N_{\max}$ of secondary particles, where the critical energy is reached. In practice, the energy of the primary particle is deposited through the atmosphere at each step of the shower development and $\Xmax$ corresponds to the atmospherical depth $X$ where $dE/dX$ is maximum, i.e., when $N_{\max}$ is reached. In the Heitler model, new particles are generated at the same time for increasing $X$. However, such an assumption is rather unphysical for different reasons. Firstly, because of the stochasticity of the process, some particles might not interact or decay at the $i-$th step of the development: therefore, they still survive at a larger atmospherical depth, causing a fluctuation in the expected number of newly created particles. Secondly, the process is dissipative, i.e., some energy is lost by developing the system between $i-$th and $i+1-$th steps. Hence, the number of particles $N(X)$ at atmospherical depth $X$ is the sum of the number $n(X)$ of newly created particles at depth $X$ and the number $n(X-dX)$, $n(X-2dX)$, etc., of surviving particles from the smaller atmospherical depths, i.e., $N(X)$ is the sum of correlated random numbers. 

Before introducing the Gumbel distribution and its interpretation for dissipative systems, it is of interest to briefly recall the standard central limit theorem (CLT). The CLT for independent and identically distributed (i.i.d.) random variables $\{x_{i}\}_{i=1,2,...,\tilde{n}}$, with finite variance, states that the statistics of their sum
\begin{eqnarray}
y_{\tilde{n}}=
\sum_{i=1}^{\tilde{n}} x_{i}
\end{eqnarray}
can be described by a Gaussian distribution in the limit $\tilde{n}\lto\infty$. More precisely, since each variable $x_{i}$ has the same finite mean $m$ and variance $\sigma^2$, the rescaled variable
\begin{eqnarray}
z_{\tilde{n}}=\frac{y_{\tilde{n}}-a_{\tilde{n}}}{b_{\tilde{n}}},
\end{eqnarray}
converges in law when $\tilde{n}\lto\infty$ to the normal distribution, if $a_{\tilde{n}}=\tilde{n}m$ and $b_{\tilde{n}}=\sigma\sqrt{\tilde{n}}$. 

However, when at least one among these conditions (i.i.d and finite second moment) is not fulfilled, the CLT is no longer valid. The case when the second moment is not finite still leads to a class of universal functions, the so-called L\'evy-stable distributions, depending on two shape parameters and two scale parameters, and the CLT can be generalized accordingly. The cases involving correlated and/or non-identically distributed random variables is much more complicated and, in general, there are no simple criteria to ensure the applicability of the CLT or its extended version (we refer to, e.g., \cite{Clusel2008} for a recent review about this topic). However, it has been recently observed that for a given class of physical systems, namely systems with dissipative interactions, the energy distribution (where the energy is defined as a sum of random variables) becomes close to a Gumbel distribution \cite{BHP,Bramwell01,Poggi01,Noullez02,Brey05,Holdsworth13}. In some very simplified cases, the Gumbel distribution has been shown to be the exact one \cite{Antal01,bertin2005global}. The Gumbel distribution initially originates from the field of extreme value statistics (see Appendix~\ref{app:EVT}), and its relation to problems of sums and to dissipative systems has been clarified recently \cite{bertin2005global,bertin2006exactly}. 

In the most simple instances, a typical stochastic dissipative system can be modeled by a one-dimensional lattice where: i) energy is injected at a boundary site with given rates, ii) energy is transferred from site $n$ to site $n+1$ and $n-1$ with known rates, and iii) energy is dissipated at site $n'$ with given rate. For specific choices of the rates, the total energy, corresponding to the sum of the energy of each site, follows the generalized Gumbel distribution \cite{bertin2005global} defined by: 
\begin{eqnarray}
\label{eq:gumbel}
\mathcal{G}(z)= \frac{1}{\sigma}\frac{\lambda^{\lambda}}{\Gamma(\lambda)} \(e^{-\lambda z-\lambda e^{-z}}\),\qquad z= \frac{x-\mu}{\sigma},
\end{eqnarray}
where $\mu$ and $\sigma$ are a location and a scale parameters, related to the mean and the spread of the distribution, respectively.
For more general choices of the rates, the exact distribution is not known, but
the Gumbel distribution is expected to remain a reasonably good approximation of the true distribution, and thus appears as a natural candidate to fit the data.
A more detailed discussion about the Gumbel distribution and its first and second moments is given in Appendix\,\ref{app:EVT}, while more details about the estimation of the parameters are given in the next section.

We argue that the generalized Heitler model, previously described, can be considered as such a dissipative system, where the quantities playing a fundamental role are re-interpreted as follows: i) the energy injection is equal to the primary energy at site $n=0$ and it is zero at the last site $n=n_{c}$, corresponding to the atmospherical depth where the critical energy is reached; ii) the energy of each site is proportional to the number of secondary particles in that site; iii) the transfer of energy coincides with the shower development from the site $n$ to its neighbor $n+1$; iv) the dissipation at site $n'$ corresponds to the missing energy per site. 

Motivated by this analogy between the two models, we can adopt the generalized Gumbel distribution to describe the statistics of the maximum number of particles in the shower development.

In the next section we use Monte Carlo simulations of EAS to show that the generalized Gumbel distribution can be used with remarkable accuracy also to describe the distribution of $\Xmax$ values corresponding to $N_{\max}$.

\section{Parameterizing the $\Xmax$ distribution of UHECR}\label{ParametrizationSection}%

For testing our model and its validity in reproducing the $\Xmax$ distribution, we have performed Monte Carlo simulations of the EAS, for different primary particle types, hadronic interaction models and energies. More specifically, the procedure used in this section follows two steps: (1.) the Gumbel density function  $\mathcal{G}(\Xmax)$ is parameterized as a function of physical observables; (2.) the goodness of the parameterized function is tested against independent simulations. It is worth mentioning that we make use of $\Xmax$ values obtained by fitting the shower profile with the Gaisser-Hillas function \cite{GH1977}.

For the first step, we have performed simulations using the \textsc{Conex} code~\cite{conex} for a large set of primary types (p, He, C, N, O, Si, Ca, Fe and photons). For nuclear primaries we also consider three different hadronic interaction models (\textsc{QGSjet}~II-03, \textsc{Sibyll}~2. and \textsc{Epos}~1.99, and their updated versions (\textsc{QGSjet}~II-04 and \textsc{Epos-LHC}), at high energies. Simulations are performed at fixed energies, from $10^{17}$ to $10^{20}$ eV in step of 0.5 (0.1) in the logarithm of E.
In order to test the goodness of our parameterization, and its extrapolation to intermediate nuclei not involved in the fit procedure, we have generated independent sets of simulations and showers initiated by nuclei (namely Li, Ne and Mn) different from those adopted to obtain the parameterization.

\begin{figure*}[!t]
\centering%
\includegraphics[width=5.2cm]{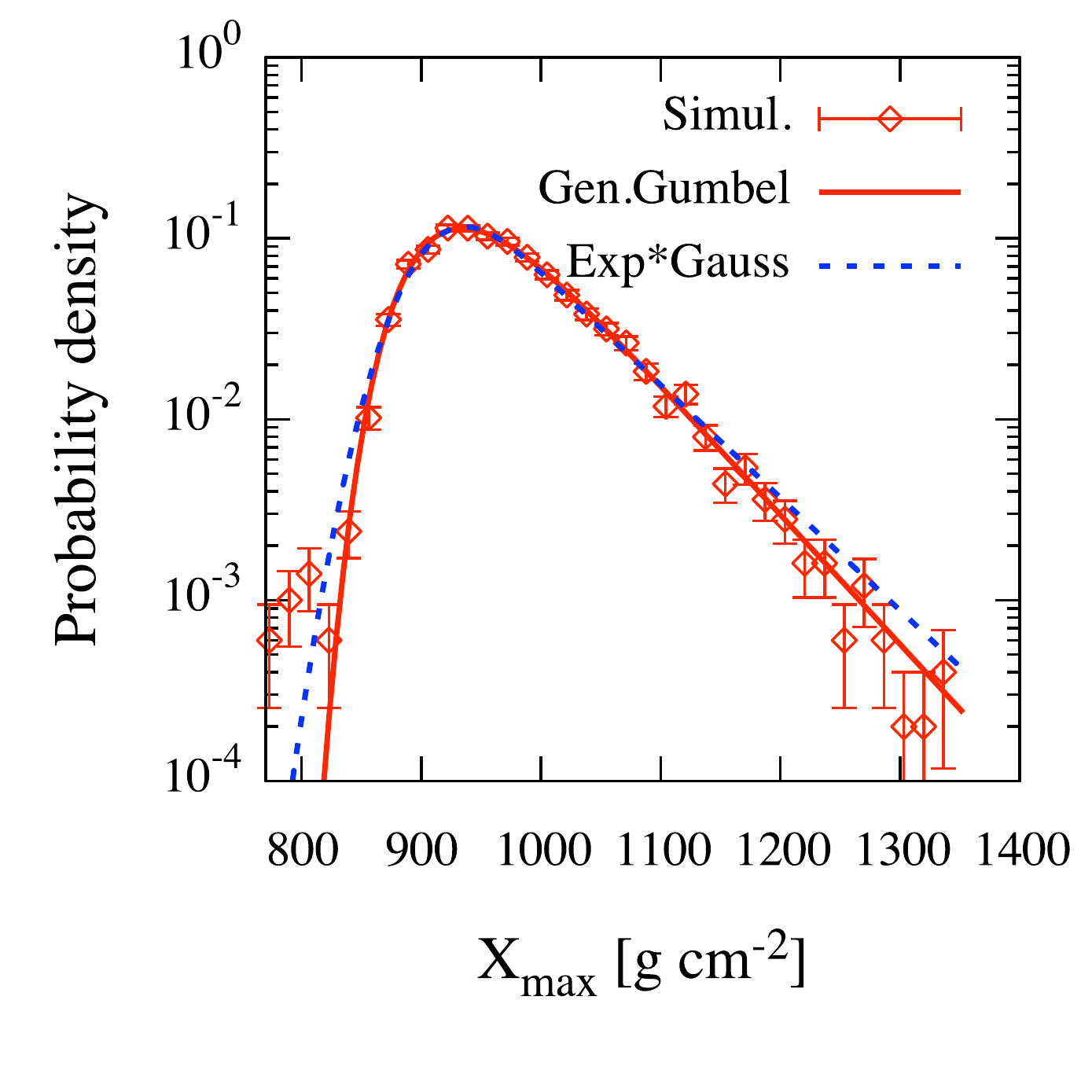}
\hspace{-3mm}
\includegraphics[width=5.2cm]{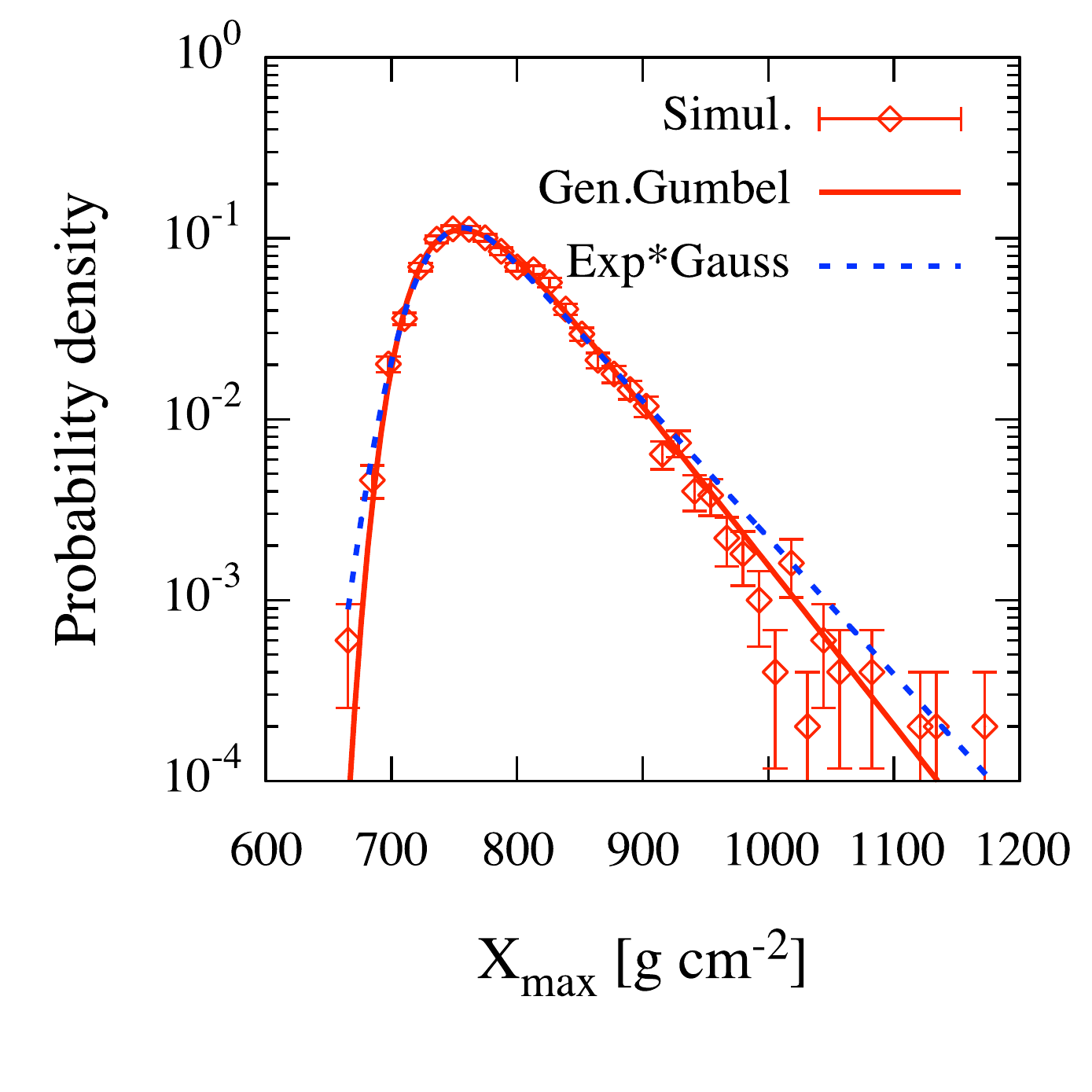}
\hspace{-3mm}
\includegraphics[width=5.2cm]{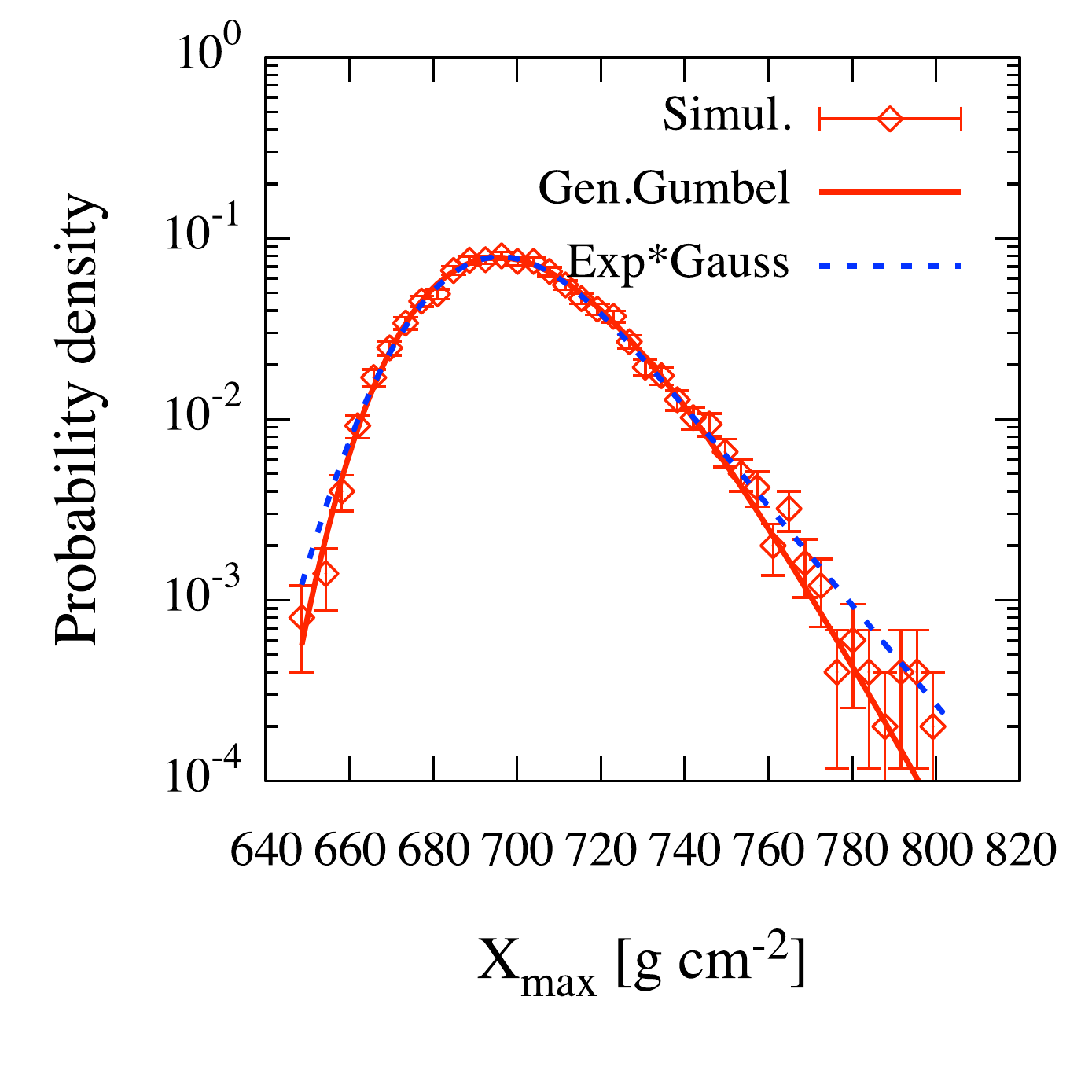}
\caption{$\Xmax$ distributions obtained from Conex simulations (markers) of photon (left panel), proton (middle) and iron (right panel) with energy 10~EeV, for the \textsc{QGSJet}~II hadronic model. The solid line indicates the best-fit obtained for the generalized Gumbel distribution, while the dashed curves indicate the best-fit obtained for the convolution between an exponential and a Gaussian distribution.}%
\label{QGSJETPlotComparison}%
\end{figure*}

As a first step we fit the generalized Gumbel distribution to the $\Xmax$ distributions obtained from simulations, for different primaries and energy intervals. We have adopted a log-likelihood maximization procedure to estimate the three parameters $\mu$, $\sigma$ and $\lambda$ (we refer to Appendix\,\ref{app:EVT} for alternative methods).
The fit was carried out for all energies and primary nuclei, and the obtained values of the parameters have been parameterized as a function of energy and nuclear mass. Some representative examples are shown in Fig.\,\ref{QGSJETPlotComparison} for the cases of photon (left panel), proton (middle panel) and iron (right panel) with energy 10~EeV and the \textsc{QGSJet}~II hadronic interaction model. The $\Xmax$ distribution obtained from Monte Carlo simulations (markers) is well reproduced by the generalized Gumbel function (red solid line). 
In Sec.~\ref{IntroductionSection} we have mentioned a model based on the convolution of an exponential function, related to the X$_{\rm{first}}$, and a Gaussian distribution, related to the fluctuations of the shower development. For comparison, we have also shown in Fig.~\ref{QGSJETPlotComparison} the best-fit curves corresponding to such an approach (blue dashed curve). Both functions provide a good description of the $\Xmax$ distribution even if the generalized Gumbel distribution better reproduces the tails for small and large values of $\Xmax$.

Thus we parameterize the density function  $\mathcal{G}(\Xmax)$ as a function of energy and primary type. However, given the significantly different development of hadronic and electromagnetic showers, the details of this parameterization are discussed below for  nuclei and photons separately.  

\subsection{Case of nuclei}\label{ParametrizationSectionNuclei}%
\begin{figure*}[!t]
\centering%
\includegraphics[width=\textwidth]{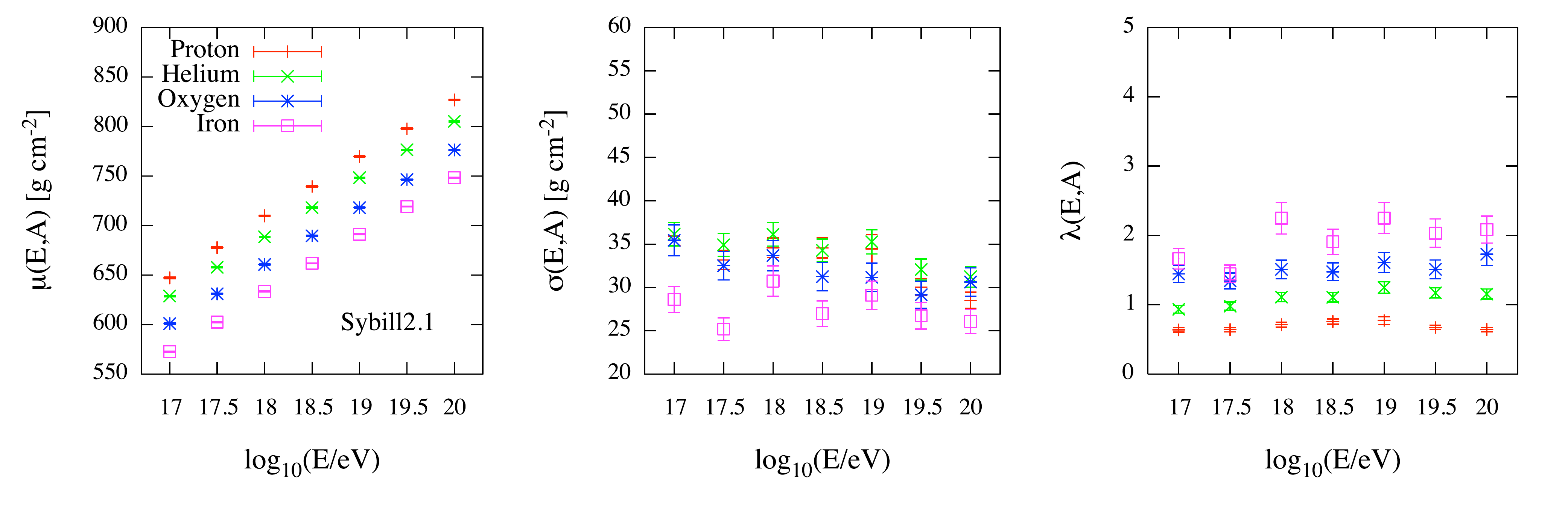}
\caption{Values of the three parameters $\mu$, $\sigma$ and $\lambda$, obtained from a likelihood fit of the generalized Gumbel distribution, as a function of energy and for a few representative examples of primary particles, for the \textsc{Sibyll}2.1 hadronic model.}%
\label{SibyllParamFitPlot}%
\end{figure*}
As introduced above, to parameterize the $\Xmax$ distribution for nuclei, we have simulated several extensive air showers in a wide range of energies $E$ and for different values of the nuclear mass $A$ and then we have parameterized the density $\mathcal{G}(\Xmax; E, A)$ as a function of $E$ and $A$. For this study we used 8 primary masses, corresponding to the p, He, C, N, O, Si, Ca and Fe. 

\begin{figure*}[!t]
\centering
\includegraphics[width=15cm]{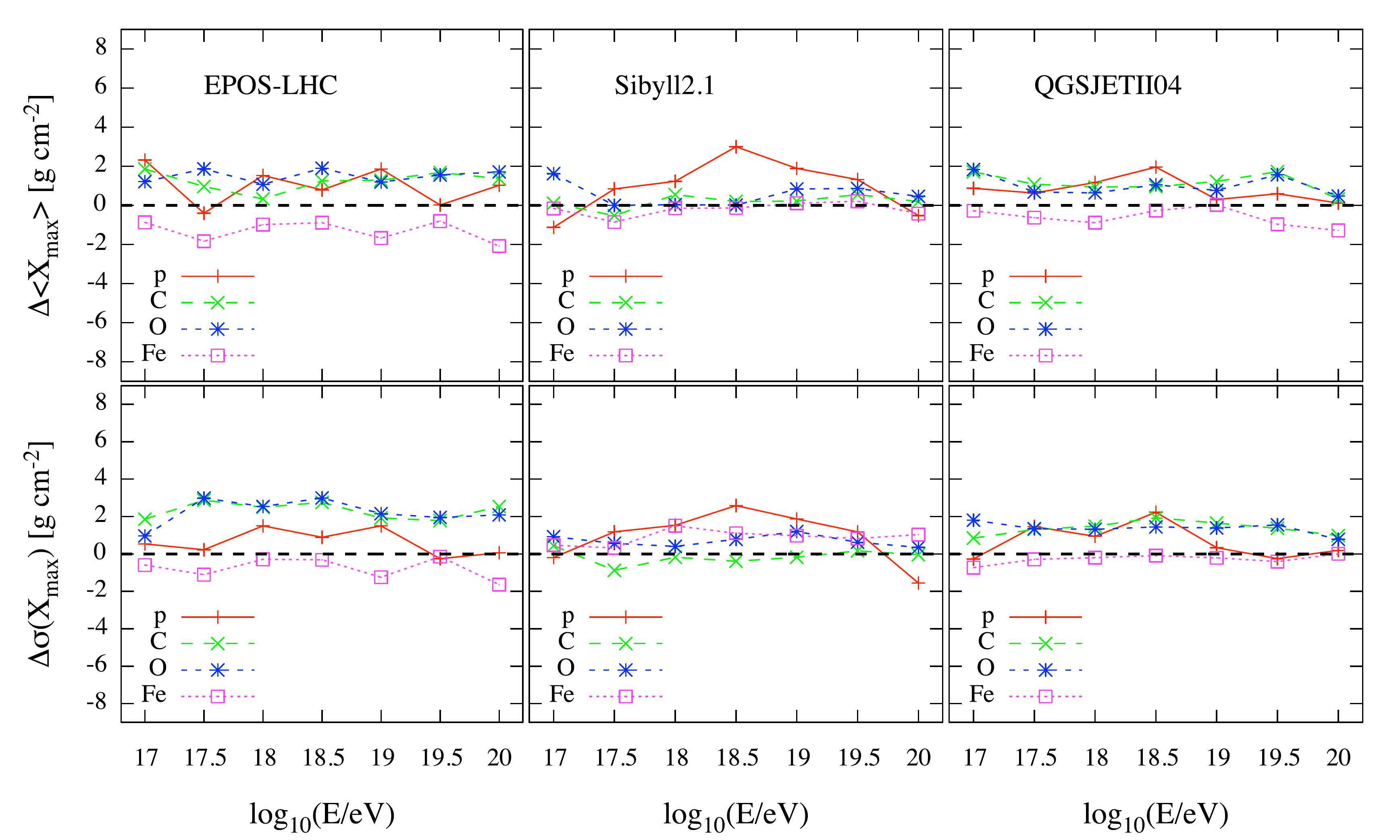}%
\caption{Mean and rms of the $\Xmax$ distribution: absolute residuals of the difference between the values obtained from generalized Gumbel expectation and the values obtained from simulations (using Gaisser-Hillas fit, i.e., Tab.\,\ref{parameterizationParamsTableNuclei}). We show the results for the three hadronic models considered in this study, namely \textsc{Epos}~LHC (left panel), \textsc{Sibyll}2.1 (middle panel) and \textsc{QGSJet}~II-04 (right panel) as a function of the energy, for a few representative examples of primary particles.}
\label{ParamResidPlot}
\end{figure*}
We have assumed the following empirical parameterizations:
\begin{eqnarray}
\label{eq:param2}\mu(A,E)&=& p_{0_{\mu}} + p_{1_{\mu}} \log_{10} (E/E_{0}) + p_{2_{\mu}} \log_{10}^{2} (E/E_{0}),\\
\label{eq:param3}\sigma(A,E)&=& p_{0_{\sigma}} + p_{1_{\sigma}} \log_{10} (E/E_{0}),\\
\label{eq:param1}\lambda(A,E)&=& p_{0_{\lambda}} + p_{1_{\lambda}} \log_{10} (E/E_{0}),
\end{eqnarray}
where $E$ and $A$ are the energy and mass of the primary particle, $E_{0}= 10^{19}$~eV is a reference value and the dependence of the parameters on the nuclear mass is empirically found as:
\begin{eqnarray}
\label{eq:param6}p_{0}^{\mu,\sigma,\lambda}&=& a_{0}^{\mu,\sigma,\lambda} + a_{1}^{\mu,\sigma,\lambda}\ln A + a_{2}^{\mu,\sigma,\lambda}\ln^{2} A,\\
\label{eq:param7}p_{1}^{\mu,\sigma,\lambda}&=& b_{0}^{\mu,\sigma,\lambda} + b_{1}^{\mu,\sigma,\lambda}\ln A + b_{2}^{\mu,\sigma,\lambda}\ln^{2} A,\\
\label{eq:param8}p_{2}^{\mu}&=& c_{0}^{\mu} + c_{1}^{\mu}\ln A + c_{2}^{\mu}\ln^{2} A.
\end{eqnarray}

For each available hadronic model, we have fitted the distribution of $\Xmax$ as a function of $\log_{10} E$ and $\ln A$. 
Therefore, a set of 21 parameters describes the full set of simulations. These parameters are summarized in Tab.\,\ref{parameterizationParamsTableNuclei} 
for three hadronic interaction models and their updated versions. It is worth noting that some experiments make use of a different $\Xmax$ estimation, determined by a quadratic 
interpolation around the maximum of the energy deposit profile. The resulting value of $\Xmax$ might differ from the one derived using a Gaisser-Hillas fit of $N(X)$ profile by a few g~cm$^{-2}$. For sake of completeness, we also provide the parameters obtained with this alternative $\Xmax$ estimation, in Appendix~\ref{appendixXmxdEdX}. However, in the following we will make use of the first parameterization by means of Tab.\,\ref{parameterizationParamsTableNuclei}.
We checked that, for each energy and primary mass, the residuals between the parameterized function and the best-fit have not any significant bias. 
In Fig.\,\ref{SibyllParamFitPlot} we show the values of the three parameters $\mu$, $\sigma$ and $\lambda$ as a function of energy, for a few representative examples of primary particles, for the case of \textsc{Sibyll}~2.1. The values of $\mu$ and $\sigma$ are related to the mean and the variance of the underlying distribution, respectively (see Appendix\,\ref{app:EVT}). As expected, the parameter $\mu$ linearly increases with energy and it decreases for increasing nuclear mass. The parameter $\sigma$ decreases for increasing nuclear mass while the parameter $\lambda$ shows the opposite trend, increasing from lighter to heavier nuclear mass. It is worth remarking that we find the same behaviour for all the hadronic models considered in this study.  

The absolute residuals of the difference between the $\langle\Xmax\rangle$ and $\sigma(\Xmax)$ values obtained from generalized Gumbel expectation and the values obtained from simulations (using Gaisser-Hillas fit, i.e., Tab.\,\ref{parameterizationParamsTableNuclei}) are plotted against energy in Fig.\,\ref{ParamResidPlot}. We show the results for \textsc{Sibyll}2.1 (middle panel) and the two up-to-date hadronic models considered in this study, namely \textsc{Epos}~LHC (left panel) and \textsc{QGSJet}~II-04 (right panel), as a function of the energy, for a few representative examples of primary particles. 
Similar results are obtained for \textsc{QGSJet}~II and \textsc{Epos}~1.99 hadronic models. The absolute residuals are smaller than about 3~gr~cm$^{-2}$ independently of energy, improving the accuracy of previous studies \cite{desouza}. It is worth remarking the goodness of our parameterization in reproducing the $\Xmax$ distribution over a wide range of nuclear masses and energy.

\begin{table}[!t]
\centering%
\begin{tabular}{|c|c|c|c|c|c|c|}%
\hline
\textsc{QGSJet}~II & $a_{0}$ & $a_{1}$ &  $a_{2}$ &  $b_{0}$ &  $b_{1}$ &  $b_{2}$\\
\hline\hline
$\mu$ & 758.444 & -10.692 & -1.253 & 48.892 & 0.02 & 0.179 \\%
$\sigma$ & 39.033 & 7.452 & -2.176 & 4.390 & -1.688 & 0.170\\%
$\lambda$ & 0.857 & 0.686 & -0.040 & 0.179 & 0.076 & -0.0130 \\ 
\hline\hline
\textsc{QGSJet}~II-04 & $a_{0}$ & $a_{1}$ &  $a_{2}$ &  $b_{0}$ &  $b_{1}$ &  $b_{2}$\\
\hline\hline
$\mu$ & 761.383 & -11.719 & -1.372 & 57.344 & -1.731 & 0.309 \\%
$\sigma$ & 35.221 & 12.335 & -2.889 & 0.307 & -1.147 & 0.271\\%
$\lambda$ & 0.673 & 0.694 & -0.007 & 0.060 & -0.019 & 0.017 \\ 
\hline\hline
\textsc{Sibyll}~2.1 & $a_{0}$ & $a_{1}$ &  $a_{2}$ &  $b_{0}$ &  $b_{1}$ &  $b_{2}$\\
\hline\hline%
$\mu$ & 770.104 & -15.873 & -0.960 & 58.668 & -0.124 & -0.023\\%
$\sigma$ & 31.717 & 1.335 & -0.601 & -1.912 & 0.007 & 0.086\\%
$\lambda$ &  0.683 & 0.278 & 0.012 & 0.008 & 0.051 & 0.003\\ 
\hline\hline
\textsc{Epos}~1.99 & $a_{0}$ & $a_{1}$ &  $a_{2}$ &  $b_{0}$ &  $b_{1}$ &  $b_{2}$\\
\hline\hline%
$\mu$ & 780.013 & -11.488 & -1.906 & 61.911 & -0.098 & 0.038\\%
$\sigma$ & 28.853 & 8.104 & -1.924 & -0.083 & -0.961 & 0.215\\%
$\lambda$ & 0.538 & 0.524 & 0.047 & 0.009 & 0.023 & 0.010\\ 
\hline\hline
\textsc{Epos-LHC} & $a_{0}$ & $a_{1}$ &  $a_{2}$ &  $b_{0}$ &  $b_{1}$ &  $b_{2}$\\
\hline\hline%
$\mu$ & 775.589 & -7.047 & -2.427 & 57.589 & -0.743 & 0.214\\%
$\sigma$ & 29.403 & 13.553 & -3.154 & 0.096 & -0.961 & 0.150\\%
$\lambda$ & 0.563 & 0.711 & 0.058 & 0.039 & 0.067 & -0.004\\ \hline%
\end{tabular}
\begin{tabular}{|c|c|c|c|}%
\hline
\textsc{QGSJet}~II & $c_{0}$ & $c_{1}$ &  $c_{2}$ \\
\hline\hline
$\mu$ & -2.346 & 0.348 & -0.086 \\%
\hline\hline
\textsc{QGSJet}~II-04 & $c_{0}$ & $c_{1}$ &  $c_{2}$ \\
\hline\hline
$\mu$ & -0.355 & 0.273 & -0.137 \\%
\hline\hline
\textsc{Sibyll}~2.1 & $c_{0}$ & $c_{1}$ &  $c_{2}$ \\
\hline\hline%
$\mu$ & -1.423 & 0.977 & -0.191 \\%
\hline\hline
\textsc{Epos}~1.99 & $c_{0}$ & $c_{1}$ &  $c_{2}$ \\
\hline\hline%
$\mu$ & -0.405 & 0.163 & -0.095 \\%
\hline\hline
\textsc{Epos-LHC} & $c_{0}$ & $c_{1}$ &  $c_{2}$ \\
\hline\hline%
$\mu$ & -0.820 & -0.169 & -0.027 \\%
\hline%
\end{tabular}

\caption{Parameters adopted for our Gumbel-based parameterization of $\Xmax$ distribution in the case of UHE nuclei.}%
\label{parameterizationParamsTableNuclei}%
\end{table}

\begin{figure}[!t]
\centering%
\includegraphics[scale=0.25]{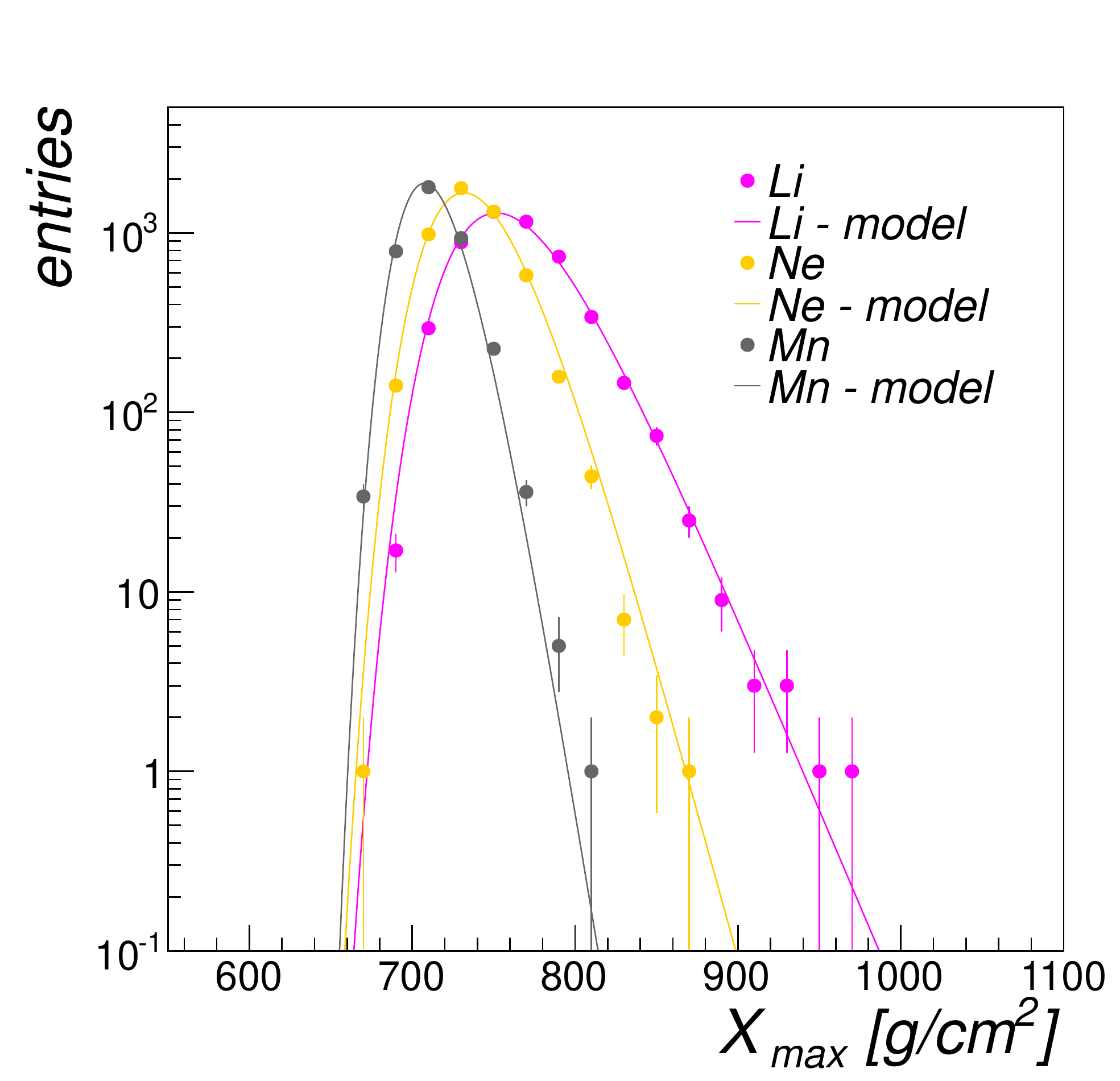}
\includegraphics[scale=0.25]{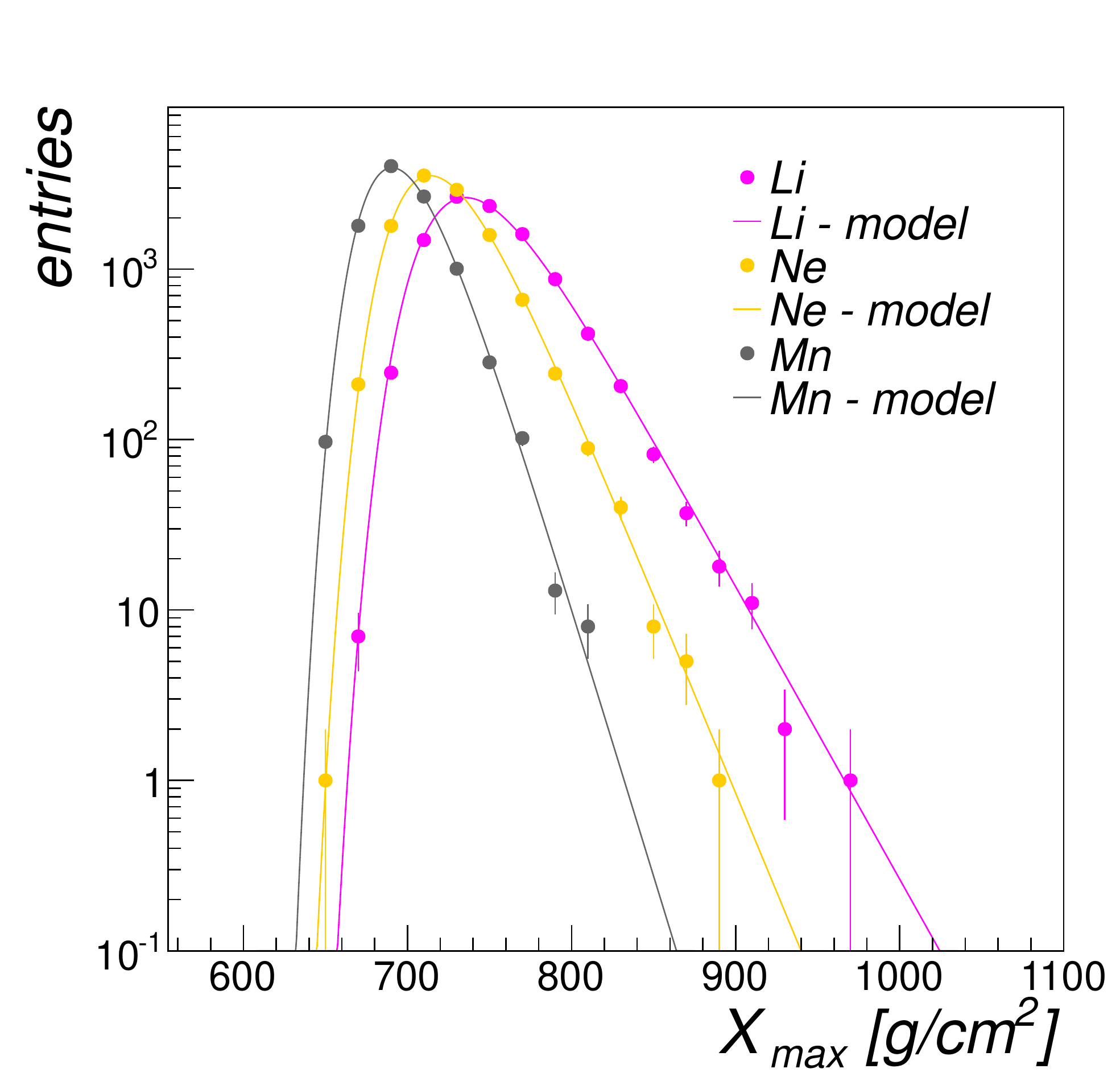}
\includegraphics[scale=0.25]{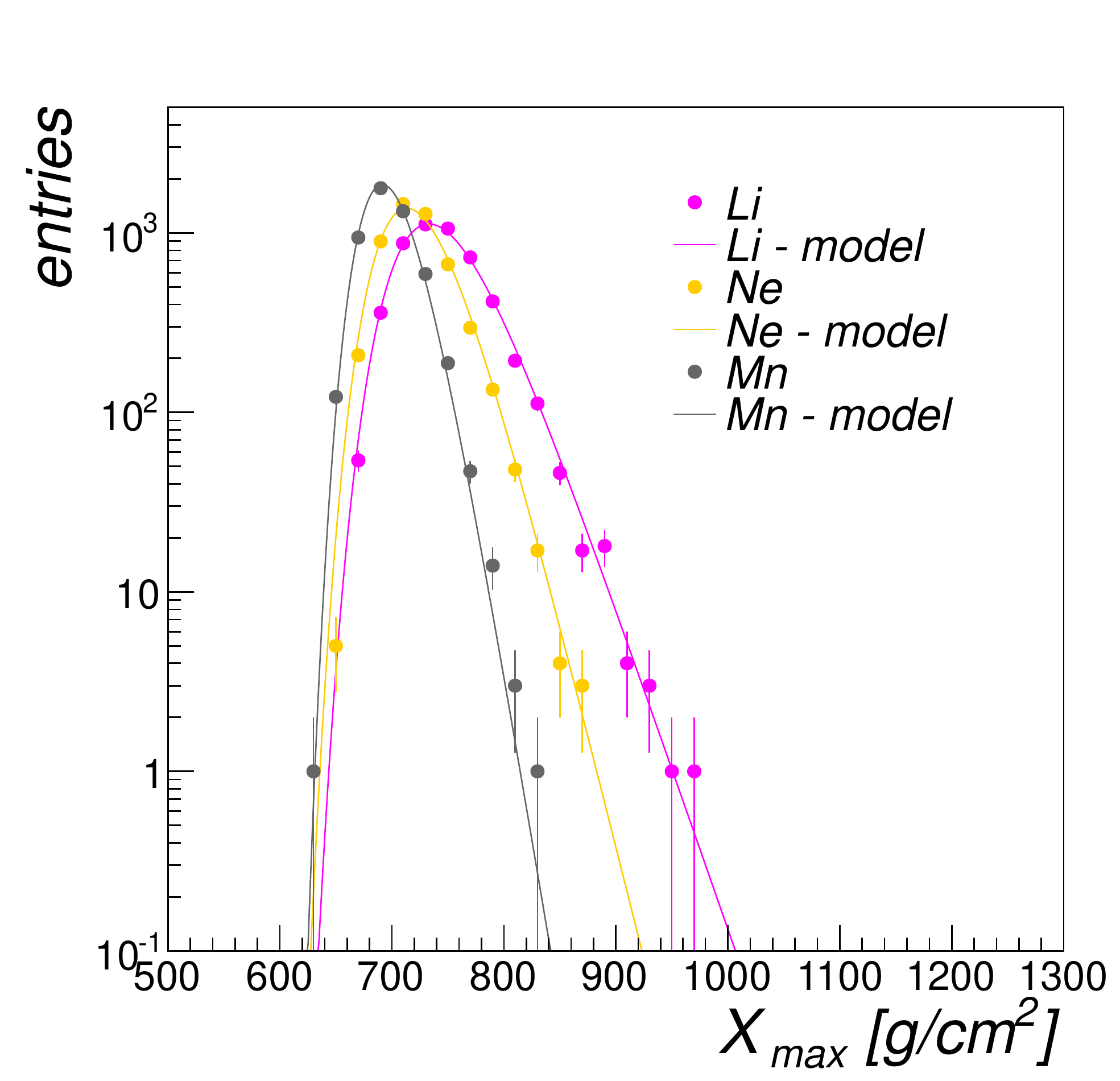}
\caption{Distribution of $\Xmax$ from simulated EAS (points) with primary energy $10^{19}$~eV initiated by Li, Ne and Mn nuclei, in the case of \textsc{Epos-LHC} (left panel), \textsc{Sibyll}2.1 (middle panel) and \textsc{QGSJet}~II-04 (right panel) hadronic models. Solid curves indicate the corresponding distributions predicted by our model.}
\label{ModelsTestFitE19Plot}
\end{figure}

Moreover we tested our model against nuclear masses not involved during the fitting procedure previously described. The results of this test are shown in Fig.\,\ref{ModelsTestFitE19Plot} where we compare the predictions of our model (lines) against the distributions of $\Xmax$ obtained from simulations (points) for the cases of Li, Ne and Mn nuclei, at $10^{19}$~eV, in the case of \textsc{Epos-LHC} (left panel), \textsc{Sibyll}2.1 (middle panel) and \textsc{QGSJet}~II-04 (right panel) hadronic models. The agreement between expectations and simulations is remarkable, even when extrapolating to nuclei not adopted to derive the parameterization.

\subsection{Case of photons}\label{ParametrizationSectionPhotons}

In this section, we extend the study done for nuclei to the case of photon induced showers. As briefly mentioned before, even if the development of electromagnetic showers can be modelled in a simpler way, several effects (Landau-Pomerachuk-Migdal, LPM~\cite{LPM, LPM2}, and pre-showering in the geomagnetic field~\cite{preshower1,preshower2,preshowerHomola}) plays an important role at high energy and can significantly modify the distribution of $\Xmax$. 
The average $\Xmax$ for photons, in the energy range considered in this paper, is typically separated from those of nuclear primaries by $\sim$~200 g~cm$^{-2}$. At energy above $\sim$~10~EeV the $\Xmax$ for photons further increases because of the LPM effect which suppresses Bremsstrahlung and pair-production cross-sections. Moreover the LPM has the effect of increasing the fluctuations in the shower development. At higher energy ($\approx$~50~EeV), photons may convert in the geomagnetic field generating a pre-shower before entering the atmosphere, and the primary energy is distributed among secondary products (e$^+$, e$^-$, photons), resulting in a smaller $\Xmax$ and in a more hadron-like behavior. However, the probability of pre-showering is a function of the local geomagnetic field, the energy and arrival direction of the primary photon. Thus a significant difference in the average $\Xmax$ may be expected in different location on the Earth.

An example of the best-fit curve of the Gumbel distribution to the $\Xmax$ for photons has been shown in Fig.~\ref{QGSJETPlotComparison} (left panel). We performed a fit of $\mathcal{G}(\Xmax)$ with $\lambda$=1 and $\mu$ and $\sigma$ as in Eq.\,(\ref{eq:m-ph}) and Eq.\,(\ref{eq:s-ph}). 

\begin{eqnarray}
\label{eq:m-ph}
\mu=a_{0} + a_1\log_{10} E + a_2\log_{10}^2E + a_3\log_{10}^3E
\end{eqnarray}

\begin{eqnarray}
\label{eq:s-ph}
\sigma = \begin{cases} a_0, & \mbox{if }E<E_0\\ a_0 + a_2\log_{10}^3(E/E_0), & \mbox{if }E>E_0\end{cases}
\end{eqnarray}
where $a_0$, $a_1$, $a_2$ and $a_3$ are the fitted parameters. Their values, obtained from the fit, are reported in Tab.\,\ref{parameterizationParamsTablePhotons}. We found that the use of a generalized Gumbel function ($\lambda\neq 1$) does not improve significantly these results while increase the number of fitted parameters.  

\begin{table}[!t]
\centering
\begin{tabular}{|c|c|c|c|c|}%
\hline
 & $a_{0}$ & $a_{1}$ & $a_{2}$ & $a_3$ \\
\hline\hline
$\mu$ & -3.16e4 & 5.50e3 & -3.16e2 & 6.14\\ 
$\sigma$  &  41.0 & 17.4 & 4.10  & \\
\hline
\end{tabular}
\caption{Parameters adopted for our Gumbel-based parameterization of $\Xmax$ distribution in the case of UHE photons.}
\label{parameterizationParamsTablePhotons}%
\end{table}

\begin{figure}[!t]
\centering%
\includegraphics[scale=0.35]{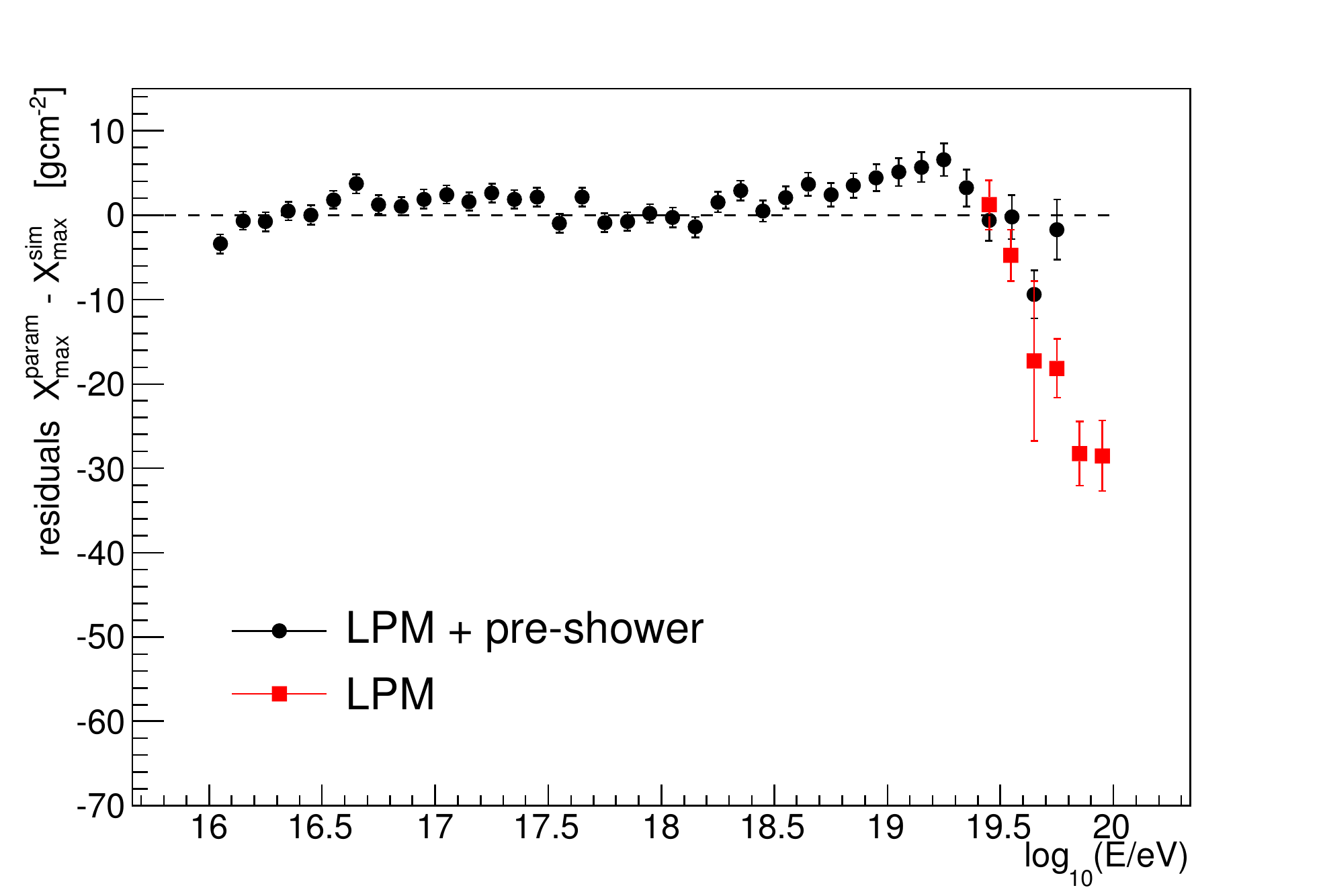}
\includegraphics[scale=0.35]{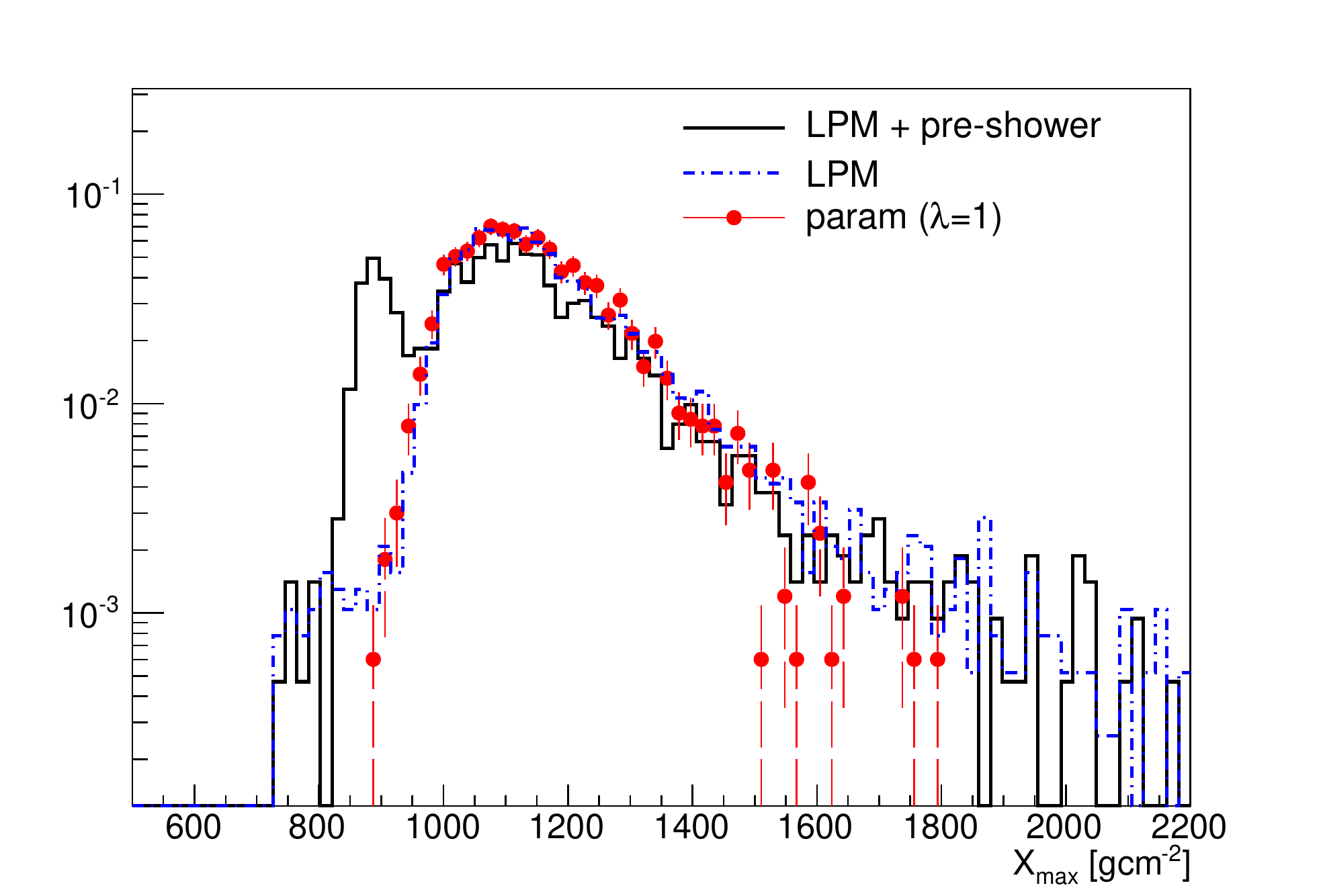}
\vspace{10pt}%
\caption{Left panel: Difference between the average values of $\Xmax$ distributions obtained from our parameterization and from simulations as a function of energy for the photon case. At higher energies the difference is also shown for the case of ``unconverted shower'' separately (red dots). Right panel: Example of the $\Xmax$ distribution at $\approx80$~EeV for the case of simulations with (black line) and without (dashed) pre-showering. The red dots indicate the result of the parameterization obtained with the classical Gumbel distribution ($\lambda$=1).}
\label{XmaxDiff}%
\end{figure}

\begin{figure*}[!t]
\centering%
\includegraphics[width=\textwidth]{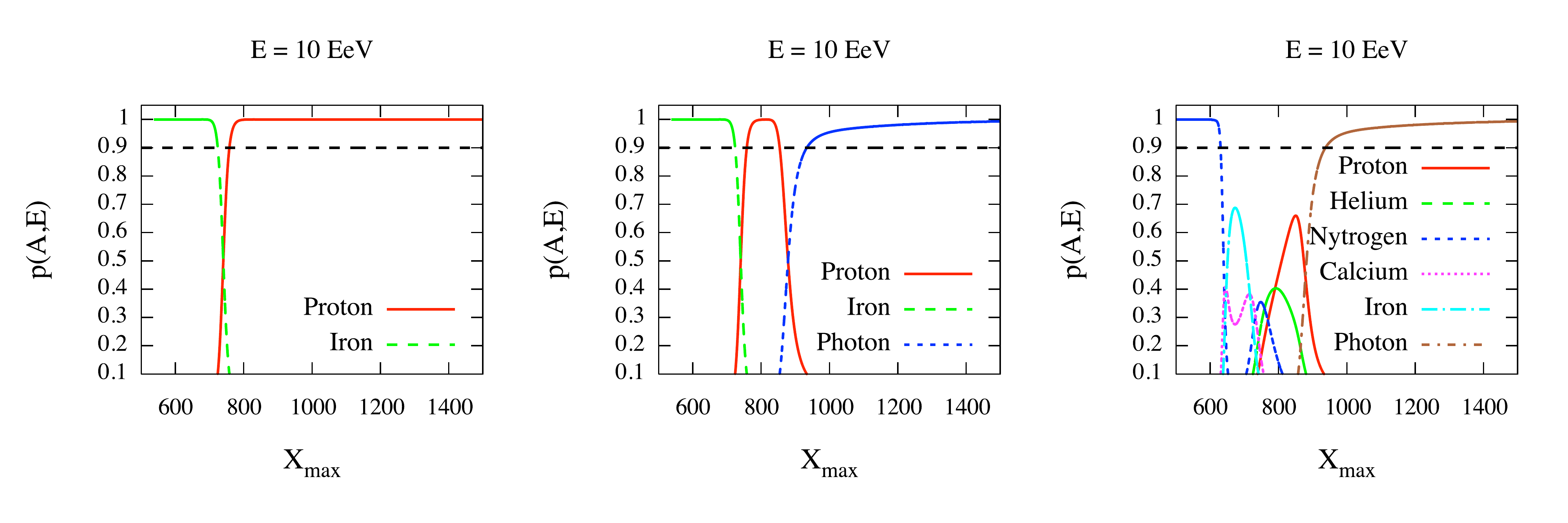}
\includegraphics[width=\textwidth]{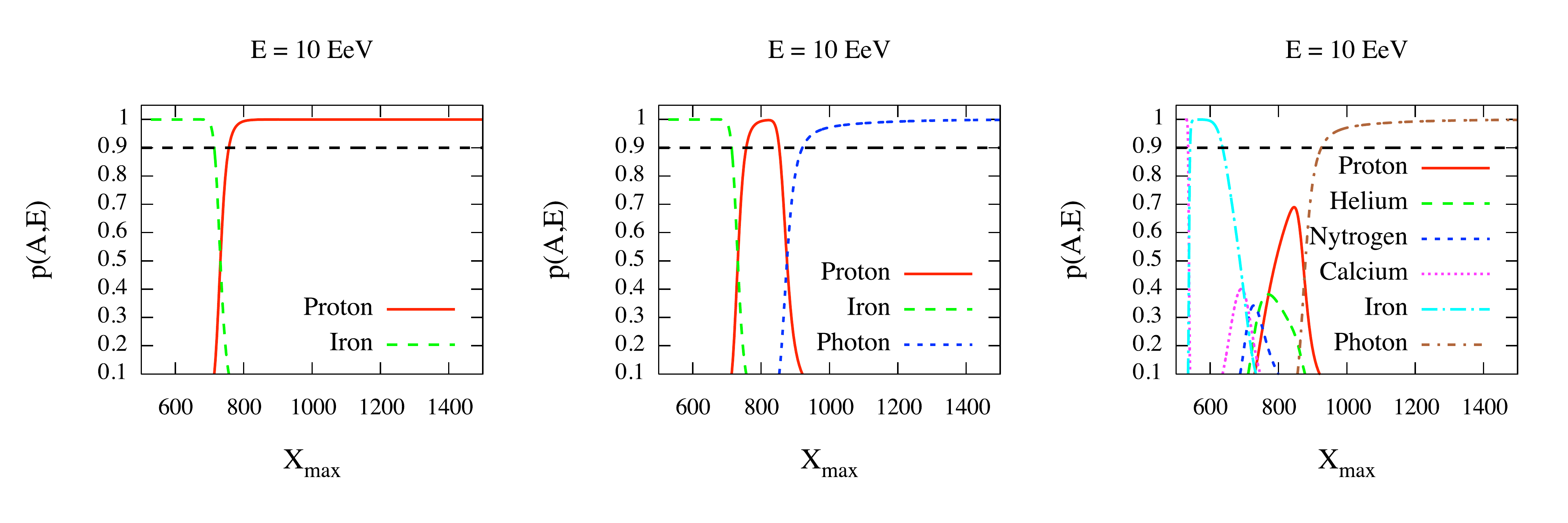}
\includegraphics[width=\textwidth]{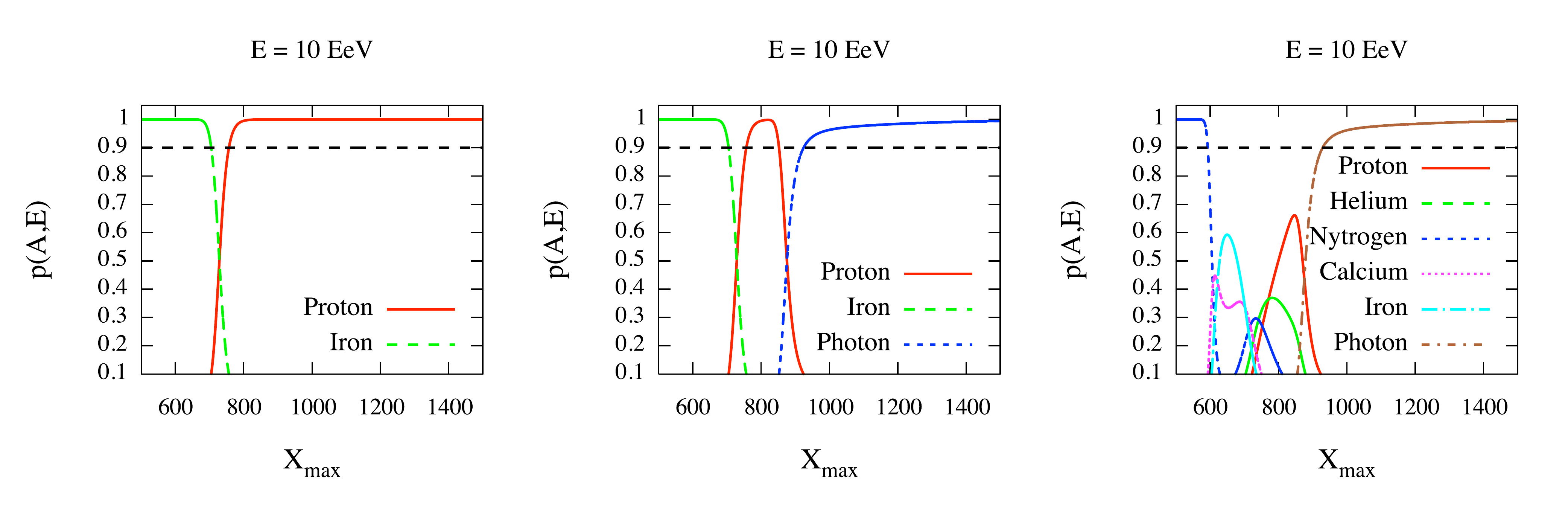}
\caption{Probability to observe an UHECR with mass $A$ as a function of $\Xmax$, in a model with 2 masses (left panels), 3 masses (central panels) and 6 masses (right panels). Expectation from three different hadronic models, namely \textsc{Epos-LHC} (top panels), Sibyll2.1 (middle panels) and \textsc{QGSJet}~II-04 (bottom panels). In each panel, the energy is fixed to $E=10$~EeV.}\label{fig:MultiSpecies}
\end{figure*}

The values of $\Xmax$  estimated from our model is in good agreement (within 5 g\,cm$^{-2}$) with the $\Xmax$ from an independent set of simulations over the full energy range, as shown in the left panel of Fig.~\ref{XmaxDiff}. As shown in figure, the method can not be extended above 50~EeV where the LPM and the pre-showering effects start to be significant.
An example of $\Xmax$ distribution at high energy (about 80~EeV) is shown in Fig.~\ref{XmaxDiff} (right panel) for the parameterized $\Xmax$ (dots) and for simulations (lines). Because of the LPM effect, the tail of the distributions at large $\Xmax$ is more enhanced while, the pre-showering is responsible of the fraction of events having small values of $\Xmax$ (around $\sim$900 g\,cm$^{-2}$). For this plot the conversion probability is the one calculated in the geomagnetic field as in the Auger South site~\cite{AugerNIM}. The distribution of $\Xmax$ for ``unconverted photons'' (dot-dashed line) is also shown for comparison.

\section{Application and Discussion}\label{DiscussionSection}

In the previous sections we have introduced the Gumbel function and we have shown that it can describe well the distribution of $\Xmax$ for a wide range of primary nuclei and for photons. We have also shown that the derived parameters are correlated to the primary energy and nuclear masses and we have briefly discussed the limitations of the method. In this section we want to briefly discuss one among the possible applications of the method introduced in the previous sections. More specifically, we propose a procedure to study the chemical composition of observed UHECR with reconstructed energy $E$ and a given measured value of $\Xmax$.

As anticipated in the introduction, several experimental results, making an extensive use of the mean and the rms of the $\Xmax$ distribution, suggest that the composition of UHECR is mixed, changing from light to a heavier component, in the energy range between 10$^{18}$~eV and 10$^{19}$~eV. 

In the following, let us assume that the observed elongation rate of UHECR, as well as the observed spectrum, is the result of the mixing of $M$ different species, with nuclear masses $A_{i}$ ($i=1,2,...,M$). For each energy $E$, we describe the $\Xmax$ distribution of each species by means of a Gumbel probability density $g(\Xmax;A_{i},E)$ (see Eq.\,(\ref{gumbeldef}) in App.\,\ref{app:EVT}) parameterized as in Sec.\,\ref{ParametrizationSectionNuclei} in the case of UHE nuclei and as in Sec.\,\ref{ParametrizationSectionPhotons} in the case of UHE photons. 

It is worth remarking that, for a fixed value of the energy $E=E^{\star}$ and a fixed value of $\Xmax = \Xmax^{\star}$, the functions $g(\Xmax^{\star};A_{i},E^{\star})$ can not be interpreted as the probability of observing a primary with mass $A_{i}$ at that energy. Our hypothesis simply consists of assuming that any observed UHECR has a mass among those one considered in the model with $M$ masses. Therefore, for any given value of $\Xmax^{\star}$ and $E^{\star}$ the sum over all species \emph{must} sum to 1, in order to obtain a meaningful physical interpretation. Hence, we define 
\begin{eqnarray}
p(A_{i};\Xmax^{\star},E^{\star})=\frac{g(\Xmax^{\star};A_{i},E^{\star})}{\sum\limits_{i=1}^{M}g(\Xmax^{\star};A_{i},E^{\star})},
\end{eqnarray}
as the probability that the observed extensive air shower has been initiated by an UHECR with reconstructed energy $E^{\star}$, measured $\Xmax^{\star}$ and mass $A_{i}$. In Fig.\,\ref{fig:MultiSpecies} we show the probability $p(A_{i};\Xmax^{\star},E^{\star})$ as a function of $\Xmax$ for $E^{\star}=10$~EeV, in a model with $M=2$ masses (left panels), $M=3$ masses (central panels) and $M=6$ masses (right panels). Moreover, we show the expectation from \textsc{Sibyll}~2.1 (middle panels) and the up-to-date hadronic models considered in this study, namely \textsc{Epos-LHC} (top panels) and \textsc{QGSJet}~II-04 (bottom panels). 

\begin{figure}[!t]
\centering%
\includegraphics[scale=0.4]{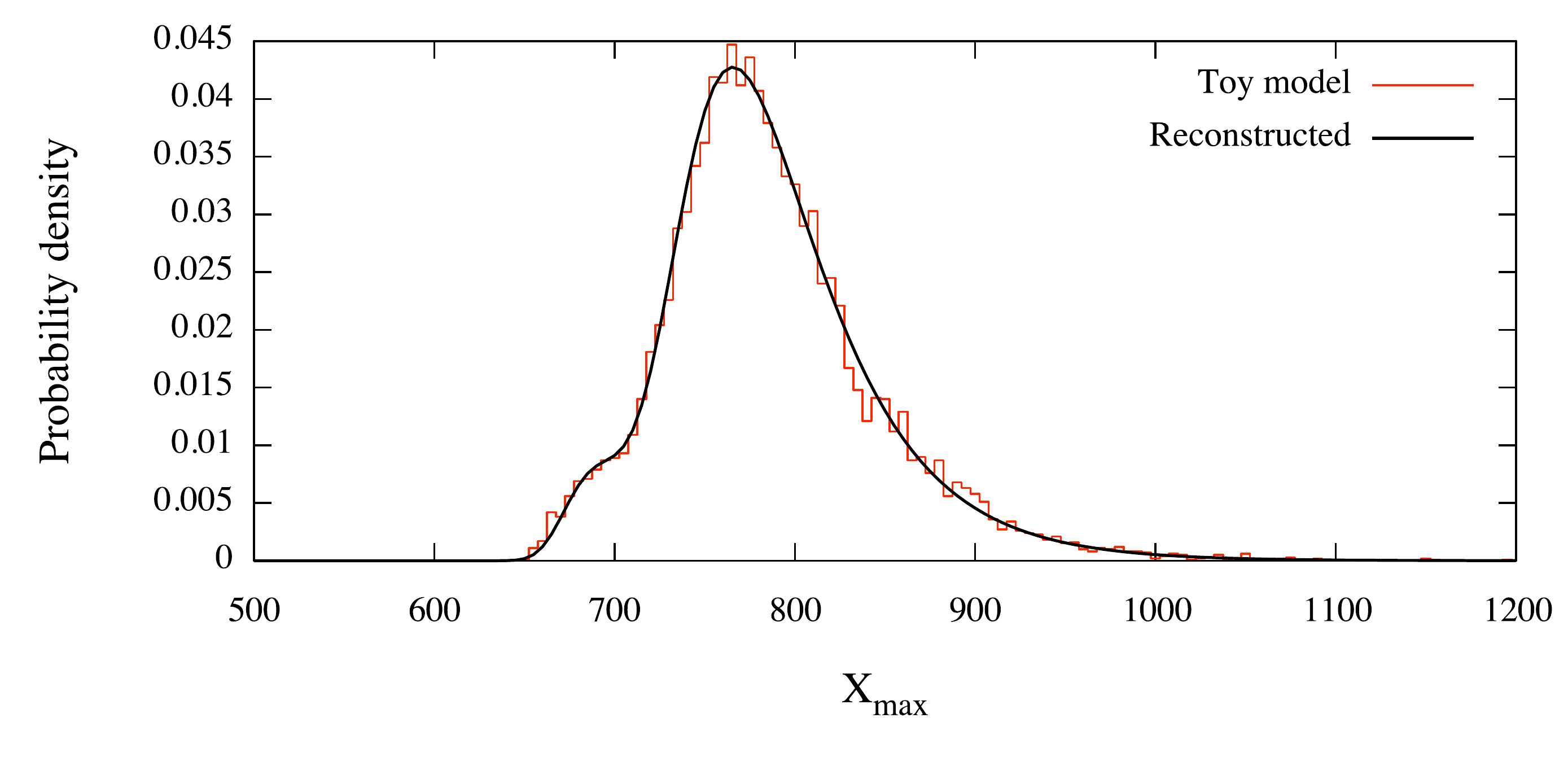}
\caption{$\Xmax$ distribution obtained from a toy model (histogram) built with $10^{4}$ values of $\Xmax$ from four nuclear masses, p, He, N and Fe, generated with \textsc{Sibyll}~2.1 at 10~EeV. In this scenario, the fractions corresponding to each nucleus are 90\%, 1\%, 1\% and 8\% respectively. The solid line indicates the reconstructed model obtained by means of the Kullback-Leibler divergence minimization.}\label{fig:FractionsKLtoy}
\end{figure}

As a practical application to real data, for instance, information shown in Fig.\,\ref{fig:MultiSpecies} can be used to estimate the fraction of observed events corresponding to each nuclear mass. In high energy physics, a standard procedure of fitting template distributions obtained from Monte Carlo simulations is given in~\cite{Barlow}. In this study, we show a very simplified application in the case of synthetic set of events, generated according to different scenarios. For each scenario we simulate $10^{3}$ random realizations of $\Xmax$ per energy bin, ranging from 1~EeV to 35~EeV. In order to create a more realistic sample of events, we smear each value of $\Xmax$ by means of a Gaussian distribution centered on $\Xmax$ with dispersion 50~g~cm$^{-2}$. More specifically, we consider two scenarios with $M=2$ nuclear masses, namely proton (p) and iron (Fe), and four scenarios with $M=4$ nuclear masses, namely proton, helium (He), nitrogen (N) and iron, where we vary the fractions corresponding to each nucleus. We choose these fractions in order to have i) scenarios dominated by lighter composition; ii) scenarios dominated by heavier composition; iii) scenarios with equally distributed composition. It is worth remarking that, for a fixed scenario, we do not vary the fractions with energy. 

Here we propose a new procedure to determine with good approximation the chemical composition of a distribution of $\Xmax$ by assigning weights $w(A_{i},E^{\star})$ ($i=1,2,...,M$) to each mass $A_{i}$, with the constraint that $\sum_{i}w(A_{i},E^{\star})=1$. The sum, opportunely normalized over all species and for any value of $\Xmax$, of the functions $g(\Xmax;A_{i},E^{\star})w(A_{i},E^{\star})$, provides a probability density $\tilde{g}(\Xmax;E^{\star},A_{1},A_{2},...,A_{M},w_{1},w_{2},...,w_{M})$ which can be used to approximate the observed $\Xmax$ distribution $p^{\star}(\Xmax;E^{\star})$ by varying the set of nuclear masses and of weights. The goodness of the approximation is estimated by means of the Kullback-Leibler divergence \cite{kullback1951information}
\begin{eqnarray}
\mathcal{D}(p^{\star}||\tilde{g})=\sum_{\Xmax} \,p^{\star}\ln \frac{p^{\star}}{\tilde{g}},
\end{eqnarray}
an information theoretical measure of the difference between two probability distributions. The Kullback-Leibler divergence is not symmetric, it is lower-bounded by 0, e.g., when $p^{\star}=\tilde{g}$, and it is not upper-bounded. Recently, the Kullback-Leibler divergence has been adopted to analyze the clustering of UHECR, by quantifying the deviation from isotropic expectation at a given angular scale \cite{dedomenico2011multiscale}. It has been shown that an alternative, although equivalent, way to perform the maximum log-likelihood fit is to minimize such a divergence (see Appendix A in \cite{dedomenico2011multiscale} and references therein for further detail). More interestingly, the Kullback-Leibler divergence has many deep theoretical interpretation, among which i) it quantifies the expected information gain about the ``true'' distribution $p^{\star}$ when using the model $\tilde{g}$; ii) it corresponds to the expected log-likelihood ratio \cite{eguchi2006interpreting}; iii) it represents the natural framework to define the Akaike Information Criterion, one of the most efficient methods to perform model selection \cite{akaike1973information,johnson2004model}.

For our purpose, thus, we are interested in minimizing the Kullback-Leibler divergence, i.e., we look for the set of mass and weight parameters providing the lowest value of $\mathcal{D}(p^{\star}||\tilde{g})$, hence, the best approximation $\tilde{g}$ to the observed distribution $p^{\star}$.

The exhaustive search over the whole parameter space is computationally intensive, hence, we make use of a faster but less precise procedure. In fact, the weights are randomly sampled from a uniform distribution between 0 and 1 and constrained to sum up to 1. A total of $10^{5}$ sets of random weights, for each mass scenario, is already sufficient for the present study. As a representative application, we show in Fig.\,\ref{fig:FractionsKLtoy} the $\Xmax$ distribution obtained from a toy model built with $10^{4}$ values of $\Xmax$ from four nuclear masses, p, He, N and Fe, generated with \textsc{Sibyll}~2.1 at 10~EeV. In this example scenario, the fractions corresponding to each nucleus are 90\%, 1\%, 1\% and 8\% respectively. The solid line indicates the reconstructed model obtained by means of the Kullback-Leibler divergence minimization, showing an excellent agreement. 

\begin{table}[!th]
\centering%
\begin{tabular}{|c|c|c|c|c|c|c|}%
\hline
\textbf{Scenario} & $f(p,E^{\star})$ & $f(Fe,E^{\star})$\\
\hline\hline
(p=90\%,Fe=10\%) & 0.87 & 0.13 \\
(p=10\%,Fe=90\%) & 0.12 & 0.88\\
(p=10\%,He=20\%,N=20\%,Fe=50\%) & 0.11 & 0.51 \\
(p=25\%,He=25\%,N=25\%,Fe=25\%) & 0.31 & 0.33 \\
(p=90\%,He=2.5\%,N=2.5\%,Fe=5\%) & 0.86 & 0.07 \\
(p=95\%,He=1\%,N=1\%,Fe=3\%) & 0.90 & 0.05\\\hline
\end{tabular}
\caption{Estimated fractions of proton and iron in datasets of events simulated with \textsc{Epos-LHC} and corresponding to different scenarios, including a Gaussian smearing of 20~g~cm$^{-2}$, in the particular case with $E^{\star}=7$~EeV, obtained by means of Kullback-Leibler divergence minimization.}
\label{tab:KLDiv}
\end{table}

We consider datasets of events simulated with \textsc{Epos-LHC} and corresponding to different scenarios, including a Gaussian smearing of 20~g~cm$^{-2}$. We consider two scenarios with $M=2$ nuclear masses (p,Fe) and four scenarios with $M=4$ nuclear masses (p,He,N,Fe). More specifically, we consider the particular case with $E^{\star}=7$~EeV, and we show the estimated fractions of proton and iron in Tab.\,\ref{tab:KLDiv} by means of Kullback-Leibler divergence minimization. By considering the presence of a smearing, the estimated fractions are in excellent agreement with the injected fractions, in each scenario, and similar performances hold for all values of the energy.

\section{Summary and outlook}\label{Section4}%

We have introduced a new method suitable to describe the distribution of shower maxima for extensive air showers initiated by ultra-high energy cosmic rays. 
Our model is motivated by an interesting relationship between the generalized Heitler model and the distribution of the sum of non-identically distributed 
variables in dissipative stochastic systems. 
In particular, we have shown that the generalized Gumbel distribution provides an excellent framework to parameterize $\Xmax$ for a wide range of 
nuclear masses, from UHE nuclei to photons, over a large energy range.

 In this paper we also provide the parameters to describe the shower maxima distribution of nuclei and photons from $10^{17}$~eV to $10^{21}$~eV, assuming 
 different hadronic interaction models, including the ones recently updated taking into account the LHC results. We tested our parameterizations with 
 independent simulations of intermediate nuclear masses finding a remarkable agreement. For photons we found that a Gumbel distribution with $\lambda=1$ provides 
 a reasonable description of the $\Xmax$ distribution in a energy regime where the LPM and the pre-showering effects are not significantly affecting the 
 shower development. Both for nuclei and photons, the resulting $\langle\Xmax\rangle$ and $\sigma(\Xmax)$ differ by less than a few g~cm$^{-2}$ from the simulated value. 

In section~\ref{DiscussionSection}, we introduced a novel method, based on information theory, to be used with our parameterization for the study of the mass composition of UHECRs. The current experimental results suggest a mixed composition from light to heavier nuclei at the highest energies. The interpretation of the experimental measurements relies on the Monte Carlo simulations of the hadronic interaction models and on the assumed scenarios of nuclear masses mixtures. Our parameterization can be used to generate in a fast way the different scenarios of masses and interaction models to reproduce the data. Moreover, since our parameterization provides analytic probability functions, it can be use to estimate, for a given scenario, the fraction of each nuclear species in the observed events. We also performed a Monte Carlo test to check the goodness of this method by generating a $\Xmax$ distribution of events for a mixed composition scenario and we found that the reconstructed fractions of nuclear masses are well in agreement with the generated ones. The extension of this approach to the experimental data is straightforward but it is out of the scope of this paper. 

\vspace{0.5truecm}\textbf{Acknowledgment.} Authors thank M. Unger for precious comments and suggestions.
M.S. acknowledges support by the BMBF Verbundforschung Astroteilchenphysik and by the Helmholtz Alliance for Astroparticle Physics (HAP).

\appendix%
\section{Extreme value theory}\label{app:EVT}

Extreme value theory is the research area dealing with the statistical analysis of the extremal values of a stochastic variable. Let $x_{i}$ ($i=1,2,...,n$) be i.i.d. random outcomes of a distribution $F$. If $M_{n}=\max\{x_{1},x_{2},...,x_{n}\}$, the probability that the maximum $M_{n}$ is less than or equal to a value $x$ is:
\begin{eqnarray}
\text{Pr}(M_{n}\leq x)=\text{Pr}(x_{1}\leq x, x_{2}\leq x, ..., x_{n}\leq x)= F^{n}(x).\nonumber
\end{eqnarray}
It can be shown that the limiting distribution $F^{n}(x)$ is degenerate and should be normalized \cite{dehaan2006extreme}. However, if there exists sequences of real constants $a_{n}>0$ and $b_{n}$ such that
\begin{eqnarray}
\text{Pr}\(\frac{M_{n}-b_{n}}{a_{n}}\leq x\)= F^{n}(a_{n}x+b_{n}),\nonumber
\end{eqnarray}
converges to a finite limit $G(x)$, 
\begin{eqnarray}
\lim_{n\lto \infty}F^{n}(a_{n}x+b_{n})=G(x)
\end{eqnarray}
then the function G(x) is the generalized extreme value (GEV) or Fisher-Tippett distribution
\begin{eqnarray}
G(z)=\left\{
\begin{array}{ll}
\exp\(-e^{-z}\) & \xi=0\\
\exp\[ -\(1-\xi z\)^{\frac{1}{\xi}} \] & \xi\neq 0
\end{array}
\right.,\quad z=\frac{x-\mu}{\sigma}
\end{eqnarray}
defined for $1-\xi z>0$ if $\xi\neq 0$ and for $z\in\mathbb{R}$ if $\xi=0$, where $\mu, \sigma$ and $\xi$ are the location, scale and shape parameters, respectively. The Gumbel distribution is related to the distribution of maxima \cite{gumbel1954statistical,gumbel2004statistics} and it is retrieved for $\xi=0$ \cite{dehaan2006extreme}. The corresponding probability density $g(x)$ is easily obtained from $G$ as
\begin{eqnarray}
\label{gumbeldef}g(x)=\frac{1}{\sigma}\exp\[ -\frac{x-\mu}{\sigma} -\exp\(\frac{x-\mu}{\sigma}\) \].
\end{eqnarray}
The two parameters $\mu$ (location) and $\sigma$ (scale) can be related to the mean $\tilde{\mu}$ and to the standard deviation $\tilde{\sigma}$ of the distribution, by means of the following relations:
\begin{eqnarray}
\tilde{\mu}&=& \mu + \gamma \sigma\\
\tilde{\sigma}^{2}&=& \frac{\pi^{2}}{6}\sigma^{2},
\end{eqnarray}
where $\gamma=0.577215...$ is the Euler constant. The generalized Gumbel distribution, dealing with the statistics of the $k$-th extremal value and the statistics of the sum of random correlated variables, includes an additional shape parameter $\lambda$, being defined by 
\begin{eqnarray}
\mathcal{G}(z)= \mathcal{G}_{0} \[e^{-z-e^{-z}}\]^{\lambda},\;\;\;\; z= \frac{x-\mu}{\sigma},
\end{eqnarray}
being $\mathcal{G}_{0}$ a normalization factor depending on $\lambda$. The case $\lambda=1$ corresponds to the classical Gumbel distribution described above. It is worth remarking here that the generalized Gumbel distribution has been recently adopted to successfully describe the statistics of many physically relevant situations, like physics of disordered systems \cite{GGD_ex1}, chemical fracture \cite{GGD_ex2}, hydrology \cite{GGD_ex3}, seismology \cite{GGD_ex4}, finance \cite{GGD_ex5,GGD_ex6} or anisotropy of ultra high energy cosmic rays \cite{dedomenico2011multiscale}, to quote only some of them.

The method of moments allows to relate the parameters of the generalized Gumbel distribution to its mean and variance. A convenient form of Eq.\,(\ref{eq:gumbel}) for fitting procedures is obtained by standardizing the samples to zero mean and unitary variance, if they exist and are finite, through the transformation $\Xmax \lto (\Xmax-\tilde{\mu})/\tilde{\sigma}$. Following Ref.\,\cite{ChapmanPaper}, it is possible to show that all parameters of the generalized Gumbel distribution in the standardized case are a function of $\lambda$:
\begin{eqnarray}
\mu= \sigma\[\psi(\lambda)-\ln(\lambda)\], \;\;\; \sigma^{2}= \frac{1}{\psi'(\lambda)},\;\;\;
\mathcal{G}_{0}= \frac{1}{\sigma}\frac{\lambda^{\lambda}}{\Gamma(\lambda)},\;\;\; \psi(\lambda)= \frac{\Gamma'(\lambda)}{\Gamma(\lambda)},
\end{eqnarray}
being $\Gamma$ the Euler gamma function and $\psi(\lambda)$ the digamma function. Such a parameterization is particularly useful in applications because only the parameter $\lambda$ has to be determined by fitting standardized data, although the parameters $\mu$ and $\sigma$ remain to be estimated.

Hence, let us define $A=\sigma^{-1}$, $B=\sigma^{-1}\psi(\lambda)$ and $C=\sigma^{-1}\[\psi^{2}(\lambda)+\psi'(\lambda)\]$. It is possible to show \cite{ChapmanPaper} that the first two non-central moments of the generalized Gumbel distribution are given by
\begin{eqnarray}
M_{1}&=&\sigma^{2}\[A\(\ln\lambda+\frac{\mu}{\sigma}\)-B\]\\
M_{2}&=&\sigma^{3}\[A\(\ln\lambda+\frac{\mu}{\sigma}\)^{2}-2\(\ln\lambda+\frac{\mu}{\sigma}\)B+C\],
\end{eqnarray}
respectively. The corresponding cumulants, i.e., the mean $\tilde{\mu}$ and the variance $\tilde{\sigma}^{2}$, are given by $\tilde{\mu}=M_{1}$ and $\tilde{\sigma}^{2}=M_{2}-M_{1}^{2}$. Finally, we obtain the relations
\begin{eqnarray}
\tilde{\mu}&=& \mu + \sigma\ln\lambda-\sigma\psi(\lambda)\\
\tilde{\sigma}^{2}&=& \sigma^{2}\psi'(\lambda),
\end{eqnarray}
completing the procedure to determine all the parameters of the generalized Gumbel distribution fitting the observation. For $\lambda=1$, an by noting that $\psi(1)=-\gamma$ and $\psi'(1)=\pi^{2}/6$, the mean and the variance of the classical Gumbel distribution are retrieved.

\section{Alternative $\Xmax$ parameterization}\label{appendixXmxdEdX}

Throughout this work we assumed the definition of $\Xmax$ as the atmospheric depth at which the profile $dN(X)/dX$ of the extensive air shower reaches its maximum $N_{max}$. As mentioned in section~\ref{ParametrizationSectionNuclei}, an alternative definition of the $\Xmax$ value, hereafter denoted as $\Xmax^{\rm{dEdX}}$, is also adopted
in literature and in shower simulation tools. $\Xmax^{\rm{dEdX}}$ is defined as the depth at which the shower energy deposit profile $dE(X)/dX$ reaches its maximum and, for instance, it
can be estimated by means of a quadratic interpolation around the maximum of $dE(X)/dX$.\\
Such a definition is mostly used by the experimentalists involved in measurements with fluorescence 
telescopes since the fluorescence light emitted along the shower development is proportional to the dE(X)/dX. \\%
We compared the values of $X_{\rm{max}}$ and $\Xmax^{\rm{dEdX}}$ obtained from our simulations for different hadronic models, primary energies and nuclei and we found a perfect linear correlation between the two variables with a systematic negative offset $\Xmax^{\rm{dEdX}}$-$X_{\rm{max}}$ ranging from -5 to -10 g/cm$^{2}$, depending on the scenario. Such differences do not alter the nature of the expected probability distribution, that is still a generalized Gumbel, while they correspondingly affect the values of the parameters (for instance $\mu$). 
It is worth mentioning that the residuals are similar to the ones shown in Fig.\,\ref{ParamResidPlot} obtained by using the Gaisser-Hillas fit. Therefore, for sake of completeness we additionally provide in Tab.\,\ref{parameterizationParamsTableNuclei2} the values of the resulting parameters when $\Xmax^{\rm{dEdX}}$ is adopted instead of $\Xmax$.

\begin{table}[!h]
\centering%
\begin{tabular}{|c|c|c|c|c|c|c|}%
\hline
\textsc{QGSJet}~II & $a_{0}$ & $a_{1}$ &  $a_{2}$ &  $b_{0}$ &  $b_{1}$ &  $b_{2}$\\
\hline\hline
$\mu$ & 756.881 & -10.982 & -1.259 & 49.665 & -0.296 & 0.251 \\%
$\sigma$ & 40.751 & 7.169 & -2.209 & 5.120 & -2.061 & 0.228\\%
$\lambda$ & 0.901 & 0.700 & -0.048 & 0.200 & 0.066 & -0.011 \\ 
\hline\hline
\textsc{QGSJet}~II-04 & $a_{0}$ & $a_{1}$ &  $a_{2}$ &  $b_{0}$ &  $b_{1}$ &  $b_{2}$\\
\hline\hline
$\mu$ & 760.023 & -12.107 & -1.364 & 57.973 & -1.836 & 0.349 \\%
$\sigma$ & 36.355 & 12.199 & -2.876 & 0.600 & -1.221 & 0.276\\%
$\lambda$ & 0.699 & 0.697 & -0.007 & 0.070 & -0.028 & 0.021\\ 
\hline\hline
\textsc{Sibyll}~2.1 & $a_{0}$ & $a_{1}$ &  $a_{2}$ &  $b_{0}$ &  $b_{1}$ &  $b_{2}$\\
\hline\hline%
$\mu$ & 768.815 & -16.440 & -0.954 & 60.039 & -0.560 & 0.044\\%
$\sigma$ & 33.472 & 0.615 & -0.535 & -1.287 & -0.242 & 0.078\\%
$\lambda$ & 0.730 & 0.267 & 0.009 & 0.029 & 0.054 & -0.009\\ 
\hline\hline
\textsc{Epos}~1.99 & $a_{0}$ & $a_{1}$ &  $a_{2}$ &  $b_{0}$ &  $b_{1}$ &  $b_{2}$\\
\hline\hline%
$\mu$ & 778.090 & -11.873 & -1.930 & 62.926 & -0.310 & 0.083\\%
$\sigma$ & 30.205 & 7.914 & -1.982 & 0.110 & -0.675 & 0.081\\%
$\lambda$ & 0.570 & 0.557 & 0.029 & 0.018 & 0.065 & -0.011\\ 
\hline\hline
\textsc{Epos-LHC} & $a_{0}$ & $a_{1}$ &  $a_{2}$ &  $b_{0}$ &  $b_{1}$ &  $b_{2}$\\
\hline\hline%
$\mu$ & 774.647 & -7.659 & -2.385 & 57.943 & -0.810 & 0.273\\%
$\sigma$ & 30.727 & 12.734 & -2.953 & 0.371 & -1.516 & 0.300\\%
$\lambda$ & 0.590 & 0.691 & 0.069 & 0.046 & 0.038 & 0.007\\ \hline%
\end{tabular}
\begin{tabular}{|c|c|c|c|}%
\hline
\textsc{QGSJet}~II & $c_{0}$ & $c_{1}$ &  $c_{2}$ \\
\hline\hline
$\mu$ & -2.222 & 0.150 & -0.058 \\%
\hline\hline
\textsc{QGSJet}~II-04 & $c_{0}$ & $c_{1}$ &  $c_{2}$ \\
\hline\hline
$\mu$ & -0.337 & 0.203 & -0.137 \\%
\hline\hline
\textsc{Sibyll}~2.1 & $c_{0}$ & $c_{1}$ &  $c_{2}$ \\
\hline\hline%
$\mu$ & -1.010 & 0.668 & -0.147 \\%
\hline\hline
\textsc{Epos}~1.99 & $c_{0}$ & $c_{1}$ &  $c_{2}$ \\
\hline\hline%
$\mu$ & -0.233 & -0.047 & -0.055 \\%
\hline\hline
\textsc{Epos-LHC} & $c_{0}$ & $c_{1}$ &  $c_{2}$ \\
\hline\hline%
$\mu$ & -1.029 & -0.157 & -0.022\\%
\hline%
\end{tabular}

\caption{Parameters adopted for our Gumbel-based parameterization of $\Xmax^{\rm{dEdX}}$ distribution in the case of UHE nuclei.}%
\label{parameterizationParamsTableNuclei2}%
\end{table}



\newpage
\begin{thebibliography}{99}

\bibitem{kampert}K.H. Kampert and M. Unger, Astropart.Phys. {\bf 35}, (2012), 660.

\bibitem{auger_mass1} J. Abraham et al. [The Pierre Auger Collaboration], Phys. Rev. Lett. {\bf 104}, (2010) 091101.

\bibitem{auger_mass2} P. Facal for the Pierre Auger Coll. ICRC 2011, arXiv:1107.4804.

\bibitem{HiRes_mass} R. Abbasi et al. [The HiRes Collaboration], Phys. Rev. Lett. {\bf 104}, (2010) 161101.

\bibitem{TA_mass} C. Jui et al. [The TA Collaboration], Proc. APS DPF Meeting, arXiv:1110.0133.

\bibitem{Yakutsk_mass} E.G. Berezhko et al., Astrop. Phys. {\bf 36}, (2012) 31.

\bibitem{WGmass2012} E. Barcikowski, J. Bellido, J. Belz, Y. Egorov, S. Knurenko, V. de Souza, , Y. Tsunesada and M. Unger for the HiRes, Pierre Auger, Telescope Array and Yakutsk Collaborations, \emph{Mass Composition Working Group Report}, UHECR 2012 Symposium, CERN 2012, EPJ Web of Conferences (in press). %

\bibitem{augerXmaxInterpretation} P. Abreu et al. [The Pierre Auger Collaboration], JCAP {\bf 1302}, (2013), 026.

\bibitem{desouza}C.J.T. Peixoto, V. de Souza and J. Bellido, (2013), arXiv:1301.5555.

\bibitem{Matthews2005} J. Matthews, Astrop. Phys. 22 (2005) 387.

\bibitem{Pierog2006} T. Pierog, R. Engel, D. Heck, Czech. J. Phys. 56 (2006), 161.%


\bibitem{bertin2005global} E. Bertin, Phys. Rev. Lett. {\bf 95} (2005), 170601.%

\bibitem{bertin2006exactly} E. Bertin, M. Clusel, J. Phys. A: Math. Gen. {\bf 39} (2006), 7607.%

\bibitem{qgsjet2} S. Ostapchenko, Phys. Rev. D {\bf 74}, (2006) 014026.%

\bibitem{sibyll} E.J. Ahn, R. Engel, T.K. Gaisser, P. Lipari, T. Stanev, Phys. Rev. D {\bf 80} (2009) 094003.

\bibitem{epos} T. Pierog, K. Werner, Phys. Rev. Lett. {\bf 101}, (2008) 171101. 
\bibitem{lhc_EAS} D. d'Enterria et al, Astrop. Phys. {\bf 35} (2011) 98.

\bibitem{qgsjet2-04} S. Ostapchenko, Phys. Rev. D {\bf 83}, (2011) 014018.%

\bibitem{epos-lhc} K. Werner, I. Karpenko, and T. Pierog, Phys. Rev. Lett. {\bf 106} (2011) 122004 


\bibitem{Heitler1954} W. Heitler, \emph{The Quantum Theory of Radiation}, third ed., Oxford University Press, London, 1954, p. 386.


\bibitem{Clusel2008} M. Clusel and E. Bertin, Int. J. Mod. Phys. B 22 (2008), 3311.%

\bibitem{BHP} S.~T. Bramwell, P.~C.~W. Holdsworth, J.-F. Pinton, Nature (London) {\bf 396}, 552 (1998).

\bibitem{Bramwell01} S.~T. Bramwell et al., Phys. Rev. E {\bf 63}, 041106 (2001).

\bibitem{Poggi01} S. Aumaitre, S. Fauve, S. mcNemara, P. Poggi, Eur. Phys. J. B {\bf 19}, 449 (2001).

\bibitem{Noullez02}
A. Noullez, J.-F. Pinton, Eur. Phys. J. B {\bf 28}, 231 (2002).

\bibitem{Brey05}
J.~J. Brey, M.~I. Garc\'{\i}a de Soria, P. Maynar, M.~J. Ruiz-Montero,
Phys. Rev. Lett. {\bf 94}, 098001 (2005).

\bibitem{Holdsworth13}
E. Bertin, P.C.W. Holdsworth (2013), arXiv:1303.4689.

\bibitem{Antal01}
T. Antal, M. Droz, G. Gy\"orgyi, Z. R\'acz, Phys. Rev. Lett. {\bf 87}, 240601 (2001).

\bibitem{GH1977} T.K. Gaisser, A.M. Hillas, Proc. 15th ICRC, Plovdiv, Bulgaria, 13-26 Aug 1977, {\bf 8} 353.%

\bibitem{conex} T. Bergmann \emph{et al.}, Astrop. Phys. {\bf 26}, (2007) 420.

\bibitem{LPM} L.D. Landau, I.Ya. Pomeranchuk, Dokl. Akad. Nauk, SSSR {\bf 92}, (1953) 535.
\bibitem{LPM2} A.B. Migdal, Phys. Rev. {\bf 103} (1956) 1811.
\bibitem{preshower1} T. Erber, Rev. Mod. Phys. {\bf 38} (1966), 626.
\bibitem{preshower2} B. McBreen and C.J. Lambert, Phys. Rev. D {\bf 24}, (1981) 2536.
\bibitem{preshowerHomola} P. Homola et al., Comput. Phys. Commun. {\bf 184}, (2013), 1468 and arXiv:1302.6939.

\bibitem{AugerNIM} J. Abraham et al. [The Pierre Auger Collaboration], Nucl. Instrum. Meth. {\bf 523} (2004) 50.
\bibitem{Barlow} R.Barlow and C.Beeston, Comp. Phys. Comm. {\bf 77}, (1993) 219.

\bibitem{kullback1951information} S. Kullback and R.A. Leibler, Ann. Math. Stat. {\bf 22} (1951) 79.
\bibitem{dedomenico2011multiscale} M. De Domenico, A. Insolia, H. Lyberis and M. Scuderi, J. Cosm. Astrop. Phys. {\bf 03}, (2011), 008.

\bibitem{eguchi2006interpreting} S. Eguchi and J. Copas, J. Multiv. Anal. {\bf 97} (2006) 2034--2040.

\bibitem{akaike1973information} H. Akaike, \emph{Information theory and an extension of the maximum likelihood principle} in 2nd International Symposium on Information Theory (Edited by B. N. Petrov and F.
Csaki) (1973), 267-281. Akademia Kiado, Budapest.
\bibitem{johnson2004model} J. B. Johnson and K. S. Omland, Trends in ecology \& evolution {\bf 19} (2004) 101--108


\bibitem{dehaan2006extreme} L. De Haan and A. Ferreira, \emph{Extreme value theory: an introduction}, Springer Verlag (2006).%
\bibitem{gumbel1954statistical} E.J. Gumbel, \emph{Statistical theory of extreme values and some practical applications: A series of lectures}, National Bureau of Standards Washington, DC (1954).%
\bibitem{gumbel2004statistics} E.J. Gumbel, \emph{Statistics of extremes}, Dover Pub. (2004).%
\bibitem{GGD_ex1} J.P. Bouchaud and M. Mezard, J. Phys. A {\bf 30}, 7997 (1997).
\bibitem{GGD_ex2} A. Baldassarri, A. Gabrielli and B. Sapoval, Europhys. Lett. {\bf 59}, (2002) 232. 
\bibitem{GGD_ex3} R. W. Katz, M. B. Parlangi and P. Naveau, Advances in Water Res. {\bf 25}, 1287 (2002).
\bibitem{GGD_ex4} D. Sornette, L. Knopoff, Y. Kagan and C. Vannest, J. Geophys. Res. {\bf 101}, 13883 (1996). 
\bibitem{GGD_ex5} F. Longin, J. Bank. Finance {\bf 24}, 1097 (2000). 
\bibitem{GGD_ex6} J.P. Bouchaud and M. Potters, \emph{Theory of Financial Risk and Derivative Pricing}, Cambridge University Press, Cambridge, 2003, 2nd edition.

\bibitem{ChapmanPaper} S.C. Chapman, G.Rowlands and N.W.Watkins, Nonl. Proc. Geoph. {\bf 9} (2002), 409.%

\end{thebibliography}
\end{document}